\documentclass[11pt]{article} 
\usepackage{times} 
\usepackage{graphicx} 

\def\bsigma{\mbox{\boldmath $\sigma$}}
\def\bxi{\mbox{\boldmath $\xi$}}
\def\E{\mbox{\rm E}}
\def\Var{\mbox{\rm Var}}
\def\ustr#1#2{\;\,\stackrel{#1}{#2}\;\,}

\begin{document}
\title{Multi-state neural networks based upon spin-glasses: a biased 
 overview}
\author{ D.~Boll\'e}

\date{}
\maketitle

Recent results are reviewed on both the time evolution and retrieval 
properties of multi-state neural networks that are based upon 
spin-glass models.
In particular, the properties of models with neuron states having  
Q-Ising symmetry are discussed for various architectures.
The main common features and differences are highlighted.

\section{Introduction}

Artificial neural networks have been widely applied to memorize and
retrieve information. During the last number of years there has been
considerable interest in neural networks with multistate neurons
(see, e.g., \cite{B00}--\cite{BVZ} and references cited therein).
Basically, such models can function as associative memories for grey-toned
or coloured patterns \cite{TI01}, \cite{IC01} and/or allow for a more 
complicated internal structure of the retrieval process, e.g., a 
distinction between the exact location and the details of a picture in 
pattern recognition and the analogous problem  in the framework of 
cognitive neuroscience \cite{SNK98}, a combination of information
retrieval based on skills and based on specific facts or data 
\cite{KI93}, \cite{S86}.

In analogy with the well-known Hopfield model \cite{Hopfielda},
\cite{Hopfieldb} the models we discuss here are built from spin-glasses
(see \cite{AGS} for the Hopfield model) with couplings defined in terms 
of embedded patterns through a learning
rule. Since one of the aims of these networks is to find back the
embedded patterns as attractors of the retrieval process, they are also
interesting from the point of view of dynamical systems.

Different types of multi-state spins (=neurons) can be distinguished 
according to the
symmetry of the interactions between the different states. The states of
the $Q$-Ising neuron can be represented by 
scalars, and the interaction between two neurons can then be written as a
function of the product of these scalars. So the $Q$- states of the neuron
can be ordered like a ladder between  a minimum and a maximum value,
usually taken to be $-1$ and $+1$.
Special cases are $Q=2$, i.e., the Hopfield model and  $Q=\infty$, i.e.,
the analogue or graded response neuron.
The states of the phasor or clock neuron can be
represented by vectors in the complex plane that are placed (equally
spaced) on the unit circle. The interaction between two neurons can then
be written as a function of the real part of the product of these
vectors indicating the state of the two neurons.
The $Q$-Potts neuron
states can be represented by $(Q-1)$ dimensional vectors that are placed
on the edges of a regular $(Q-1)$ dimensional simplex and the interaction
between two neurons is then a function of the scalar product of these
vectors, which is either $(Q-1)/Q$ or $1/Q$.
For $Q=2$ the three types of neurons are the same after proper
rescaling, and for $Q=3$ the phasor and the Potts neurons are
equivalent.

The neural network models built with these multi-state neurons have an
immediate analogon in spin-glass systems (cfr.,
e.g., \cite{SK} and \cite{GS}, respectively,  \cite{NS86}, \cite{ES}).
Of course, these  types of multi-state neurons do not
exhaust all possibilities for constructing models. We also mention
the recently considered Ashkin-Teller and  Blume-Emery-Griffiths neural
network models that are based again upon their spin-glass counterparts
(see, e.g., \cite{NS93}, respectively, \cite{ACN00} and references 
therein),
because, as we will argue, they are especially relevant for modeling
more sophisticated features of real biological networks and/or from an
information theoretic point of view.

Besides these neuron states one also needs to specify
an architecture indicating how the neurons are connected with each other.
Several architectures have been studied in the literature for different
purposes. From a practical applications point of view mostly perceptrons
or, more general, layered feedforward networks are used since a very
long time. Fully connected attractor networks with symmetric couplings,
like the  Hopfield model, satisfy the detailed balance principle and
hence a Hamiltonian can be defined. The behaviour of such a network can 
then be studied by focusing on this Hamiltonian. An important feature 
of these attractor networks is the occurrence of feedback \cite{BKS}. 
Diluted architectures where only a fraction of the neurons are
connected, are relevant both from the biological point of view and to 
model the breakdown of synaptic couplings causing loss of information.
In particular, symmetrically extremely diluted models still allow for 
a Hamiltonian description but some feedback survives. Asymmetrically 
extremely diluted models are considered because their dynamics can be 
solved exactly since there are no feedback correlations.

Finally, one needs to give an explicit learning rule for the couplings
(e.g, Hebb \cite{Hebb}, pseudo-inverse \cite{PI}) or a strategy to find
the couplings giving the best performance (Gardner method 
\cite{Ga}, \cite{GD}).

For a more complete overview of the field from a physics point of view
we refer to the textbooks \cite{Amit}--\cite{EV}, and to  
\cite{DH91}--\cite{Sa}.

Here we review some of the most recent results on multi-state neural
networks. In particular, we focus on the models with $Q$-Ising symmetry, 
i.e., the $Q$-Ising network mainly with $Q=3$, and the
Blume-Emery-Griffiths network. Both the dynamical time evolution and the 
thermodynamic and retrieval properties for these models with various
architectures and a Hebb-type learning rule are discussed. 

The methods used are standard by now but have to be slightly extended to
accomodate the multi-state character of the neuron. First, in order to 
study the time evolution under parallel updates of the neurons we mainly 
use the signal-to-noise analysis. There exist different versions of this
method  in the literature, e.g., \cite{K}--\cite{SFb} (see 
\cite{Nishimori} for further references), \cite{PZFC1}, \cite{PZFC2}. 

In more detail, splitting the local field of the model in a signal part
from the condensed patterns and a noise part from the rest of the
patterns, and employing systematically the
law of large numbers (LLN) and the central limit theorem (CLT) we derive
the evolution of the distribution of the local field at every time step.
This allows us to obtain a recursive scheme for the evolution of the
relevant order parameters in the system.
The details of this approach depend in an essential way on the
architecture because different temporal correlations are possible.

For extremely diluted asymmetric \cite{DGZ}--\cite{BDEKT} (and references 
therein) and layered feedforward architectures \cite{DKM}--\cite{BET}
 (and references therein) recursion relations are obtained in closed 
form directly for the relevant order parameters.
This has been possible because in these types of networks there
are no feedback correlations as time progresses. As a technical
consequence the local field contains only Gaussian noise leading to an
explicit solution.

For the parallel dynamics of networks with symmetric connections,
however, things are quite different  \cite{BJSF}, \cite{PZFC1},
\cite{PZFC2}, \cite{MHV}--\cite{BBSV} (and references therein).
Even for extremely diluted  versions of these systems \cite{BJSsymdil}--
\cite{TE01} (and references therein)
feedback correlations become essential from the second time step onwards,
complicating  the dynamics in a nontrivial way.
Therefore, explicit results concerning the time evolution of the order
parameters for these models have to be obtained indirectly by starting
from  the distribution of the local field.
Technically speaking, both for the symmetrically diluted and fully
connected  architectures the local field contains both a discrete and a
normally distributed part. In both cases this discrete part prevents
a closed-form solution of the dynamics for the relevant order
parameters. Nevertheless,  the development of a recursive scheme is
possible in order to calculate their  time evolution.

By requiring through these recursion relations that the local field
becomes time-independent implying that most of the discrete noise part
is neglected, we can obtain stationary equations for the order parameters.

Since no closed-form solution of this dynamics is possible and the
results are technically complicated, it is
worthwhile to apply an alternative method, the generating functional
approach \cite{SMR73}, \cite{D78} (for a recent review see, e.g.,  
\cite{Coolen} and references therein) to solve this feedback
dynamics. This approach enables one to find all relevant physical order
parameters at any time step via the derivation of the generating
functional. Comparing this approach with the signal-to-noise ratio
analysis and with numerical simulations it turns out \cite{BBSV} that 
beyond the third time step of the dynamics the signal-to-noise analysis, 
as applied in the literature mentioned above is not completely correct for
those parameters of the system corresponding to spin-glass behaviour.
The full details of this, showing that a technical assumption concerning
the feedback correlations is not valid, although it has little
effect in most of the retrieval region of the networks, are worked out 
first for the simpler $Q=2$ Hopfield model \cite{BBVp} and are beyond
the scope of the present overview.

Secondly, the fixed-point equations for the symmetric models, which are 
governed by an Hamiltonian, can also be derived using thermodynamic replica
mean-field theory \cite{AGS}, \cite{MPV}. This allows us to write down an
expression for the free energy and obtain from it fixed-point equations 
for the order parameters. Thermodynamic and retrieval properties, e.g., 
the maximal storage capacity, can be discussed through the appropriate 
phase diagrams. Most results in the multi-state literature treat models
with sequential updating in the replica-symmetric approximation, e.g., 
 \cite{R}--\cite{BCS00}  (and references therein)  for the 
$Q$-Ising and  \cite{BV03} for the  Blume-Emery-Griffiths model. 
These works use Hebb-type learning rules for the couplings. To obtain the
optimal storage capacity by finding the optimal couplings which give the 
best  performance of the network for a specific set of patterns, the 
Gardner-approach can be used to these models. This method treats
the couplings as dynamical variables and by using a replica analysis the
minimal volume fraction of coupling space is calculated ensuring that 
this specific set of
patterns can still be embedded in the network with a certain basin of
attraction. Results for multi-state networks with $Q$-Ising type neurons 
can be found, e.g., in \cite{MKB}--\cite{BPS} (and references therein). 

We remark that the Ashkin-Teller neural network briefly mentioned above
will not be discussed here. For recent results on this model and its
relation to other networks we refer to \cite{BK02}, \cite{BK99} and to
\cite{Kthes}.

The rest of this contribution is organized as follows. In Sections 2 and
3 we consider the $Q$-Ising model, respectively the
Blume-Emery-Griffiths (BEG) model. Each Section is
divided in 3 subsections. Subsection 1 defines the model, its dynamics, 
its relevant order parameters and its measures for the retrieval quality.
In subsection 2 we use the signal-to-noise analysis in order to derive 
a recursive scheme for the evolution of the distribution of the local 
field, leading to recursion relations for the order parameters. The 
differences between the various architectures are outlined. 
Subsection 3 discusses the statics of the model describing the phase
diagrams, focusing on the retrieval properties. In Section 4 we briefly
describe the results of a Gardner approach to these models.  Finally,  
a short conclusion is given in Section 5.

This review is limited in both scope and length so that some details
and/or contributions could not be mentioned. They are referred to, 
directly or indirectly, in the references.

\section{$Q$-Ising neural networks}
\subsection{The model} \label{sec:mod}

Consider a neural network  consisting of $N$ neurons which can take
values $\sigma_i, i=1, \ldots ,N$ from a discrete set
        $ {\cal S} = \lbrace -1 = s_1 < s_2 < \ldots < s_Q
                = +1 \rbrace $.
The $p$ patterns to be stored in this network are supposed to
be a collection of independent and identically distributed random
variables (i.i.d.r.v.), $\{{\xi}_i^\mu \in {\cal S}\}$,
$\mu =\{1,\ldots,p\}$,
with zero mean, $E[\xi_i^\mu]=0$, and variance $A=\Var[\xi_i^\mu]$. The
latter is a measure for the activity of the patterns. We remark that for
simplicity we have taken the patterns and the neurons out of the same
set of variables but this is no essential restriction.
Given the configuration
        ${\bsigma}_N\equiv\{\sigma_j(t)\},
        j=\{1,\ldots,N\}$,
the local field in neuron $i$ equals
\begin{equation}
        \label{eq:h}
        h_i({\bsigma}_{N}(t))=
                \sum_{j=1}^N J_{ij}(t)\sigma_j(t)
\end{equation}
with $J_{ij}$ the synaptic coupling from neuron $j$ to neuron $i$.
In the sequel we write the shorthand notation $h_{N,i}(t) \equiv
h_i({\bsigma}_{N}(t))$.

It is clear that the $J_{ij}$ explicitly depend on the architecture.
For the extremely diluted (ED), both symmetric (SED) and asymmetric
(AED), and
the fully connected (FC) architectures the  couplings are
time-independent and the diagonal terms are absent, i.e. $J_{ii}=0$. The
configuration  ${\bsigma}_{N}(t=0)$ is chosen as input.
For the layered feedforward (LF) model the time dependence of the
couplings is relevant because the set-up of the model is somewhat
different. There, each neuron in layer $t$ is unidirectionally 
connected to all neurons on layer $t+1$ and $J_{ij}(t)$ is the strength
of the coupling from neuron
$j$ on layer $t$ to neuron $i$ on layer $t+1$. The state
${\bsigma}_{N}(t+1)$ of layer $t+1$ is determined by the state
${\bsigma}_{N}(t)$ of the previous layer $t$.

In all cases the couplings are chosen according to the Hebb rule such
that we can write
\begin{eqnarray}
     J_{ij}^{ED}&=&\frac{c_{ij}}{CA}
               \sum_{\mu =1}^p \xi_i^\mu \xi_j^\mu
        \quad \mbox{for} \quad i \not=j       \,,
        \label{eq:JED}  \\
     J_{ij}^{FC}&=&\frac{1}{NA}
               \sum_{\mu =1}^p \xi_i^\mu \xi_j^\mu
        \quad \mbox{for} \quad i \not=j       \,,
        \label{eq:JFC}  \\
    J_{ij}^{LF}(t)&=&\frac{1}{NA}
               \sum_{\mu =1}^p \xi_i^\mu(t+1)\, \xi_j^\mu(t)
              \,,
        \label{eq:JLF}
\end{eqnarray}
with the $\{c_{ij}=0,1\}, i,j =1, \ldots,N $ chosen to be i.i.d.r.v. with
distribution
$\mbox{Pr}\{c_{ij}=x\}=(1-C/N)\delta_{x,0} + (C/N) \delta_{x,1}$ and
satisfying for symmetric dilution $c_{ij}=c_{ji},\,\,\,c_{ii}=0 $, and
for asymmetric dilution that $c_{ij}$ and $c_{ji}$ are  statistically 
independent (with $c_{ii}=0$).

All neurons are updated in parallel through the spin-flip dynamics
defined by the transition probabilities
\begin{equation}
      \Pr \{\sigma_i(t+1) = s_k \in {\cal S} | \bsigma_{N}(t) \}
        =
        \frac
        {\exp [- \beta \epsilon_i(s_k|\bsigma_{N }(t))]}
        {\sum_{s \in {\cal S}} \exp [- \beta \epsilon_i
                                   (s|\bsigma_{N }(t))]}\,.
\label{eq:trans}
\end{equation}
Here the energy potential $\epsilon_i[s|{\bsigma}_{N}]$
is defined by \cite{R}
\begin{equation}
        \epsilon_i[s|{\bsigma}_{N}]=
                -\frac{1}{2}[h_i({\bsigma}_{N})s-bs^2]
         \,, \label{eq:energy}
\end{equation}
where $b>0$ is the gain parameter of the system.
The zero temperature limit $T=\beta^{-1} \rightarrow 0$ of this dynamics
is given by the updating rule
\begin{equation}
        \label{eq:enpot}
        \sigma_i(t)\rightarrow\sigma_i(t+1)=s_k:
                \min_{s\in{\cal S}} \epsilon_i[s|{\bsigma}_{N}(t)]
            =\epsilon_i[s_k|{\bsigma}_{N }(t)]
\,.
\end{equation}
This updating rule (\ref{eq:enpot}) is equivalent to using a gain 
function $\mbox{g}_b(\cdot)$,
\begin{eqnarray}
        \label{eq:gain}
        \sigma_i(t+1) &  =   &
               \mbox{g}_b(h_{N,i}(t))
                  \nonumber      \\
               \mbox{g}_b(x) &\equiv& \sum_{k=1}^Qs_k
                        \left[\theta\left[b(s_{k+1}+s_k)-x\right]-
                              \theta\left[b(s_k+s_{k-1})-x\right]
                        \right]
\end{eqnarray}
with $s_0\equiv -\infty$ and $s_{Q+1}\equiv +\infty$. For finite $Q$,
this gain function $\mbox{g}_b(\cdot)$ is a step function.
The gain parameter $b$ controls the average slope of $\mbox{g}_b(\cdot)$.

In order to measure the retrieval quality of the system one can use the
Hamming
distance between a stored pattern and the microscopic state of the network
\begin{equation}
        d({\bxi}^\mu,{\bsigma}_N(t))\equiv
                \frac{1}{N}
                \sum_{i}[\xi_i^\mu-\sigma_i(t)]^2         \,.
\end{equation}
This  introduces the main overlap and the arithmetic mean of the
neuron activities
\begin{equation}
        \label{eq:mdef}
        m_N^\mu(t)=\frac{1}{NA}
                \sum_{i}\xi_i^\mu \sigma_i(t),
                \quad \mu =1, \ldots, p\, ; \quad
        a_N(t)=\frac{1}{N}\sum_{i}[\sigma_i(t)]^2    \,.
\end{equation}
We remark that for $Q=2$ the variance of the patterns $A=1$, and the
neuron activity $a(t)=1$. For the LF architecture we recall that
 $\xi_i^\mu$ depends on $t$.

In this overview we mainly consider the patterns to be uniformly 
distributed
(e.g., $A=2/3$ for $Q=3$). For low-activity networks ($A$ small, e.g., 
$A<<2/3$ for $Q=3$) a better measure for the retrieval quality is the
mutual information. We refer to the literature for a further discussion
of this point \cite{Okada1996}--\cite{BD00} (and references therein).

\subsection{Solving the dynamics} \label{sec:gensch}
\subsubsection{Correlations}

We first discuss some of the geometric properties of the various
architectures which are particularly relevant for the understanding of
their long-time dynamic behaviour.

For a FC architecture there are two main sources of strong
correlations between the neurons complicating the dynamical evolution~:
feedback loops and the common
ancestor problem \cite{BKS}. Feedback loops occur when in the
course of the time evolution, e.g., the following string of connections is
possible: $i \rightarrow j \rightarrow k \rightarrow i$. We remark that
architectures with symmetric connections always have these feedback
loops. In the absence of these loops the network functions in fact as a
layered  system, i.e., only feedforward connections are possible.
But in this layered architecture common ancestors are still present when,
e.g., for  the sites $i$ and $j$ there are sites in the foregoing time
steps that have a connection with both $i$ and  $j$.

In AED architectures these sources of correlations are absent.
This  class of neural networks was introduced in connection with
$Q=2$-Ising  models \cite{DGZ}. We recall that the couplings are
then
given by eq.~(\ref{eq:JED}) and that in the limit $N \rightarrow \infty$
two important properties of this network are essential
\cite{DGZ}, \cite{KZ}.
The first property is the high asymmetry of the connections, viz.
\begin{equation}
   \Pr\{c_{ij} = c_{ji}\} = \left(\frac{C}{N}\right)^2 \, , \quad
        \Pr\{c_{ij} = 1 \wedge c_{ji} = 0\} = \frac{C}{N}
            \left(1-\frac{C}{N}\right).
            \label{eq:G2}
\end{equation}
Therefore, almost all connections of the graph
$G_{{ N}}({\bf c}) = \{(i,j) : c_{ij}=1, i,j \neq i=1,\ldots,N \}$
are  directed~: $c_{ij} \neq c_{ji}$.
The second property in the limit of extreme dilution is the directed local
Cayley-tree
structure of the graph $G_{{ N}}({\bf c})$. By the arguments above the
probability  that $k$ connections are directed towards a
given site $i$ becomes a Poisson distribution in the limit of
extreme dilution and the mean value of the number of in(out) connections
for this site $i$ is $C$. The probability
that the sites $i$ and $i'$ have site $j$ as a common ancester is
obviously $C/N$, hence the probability that the sites $i$ and $i'$ have
disjoint clusters of ancestors
approaches $(1-{C^t}/{N})^{C^t} \simeq \exp (-{C^{2t}}/{N})$
for $N \gg 1$.

So we find that in the limit of extreme dilution 
almost all (i.e. with probability 1) feedback loops  are eliminated.
and any finite number of neurons have almost all disjoint clusters
of ancestors.
So we first dilute the system by taking $N \rightarrow \infty$ and then
we take the limit $C \rightarrow \infty$ in order to get infinite
average connectivity allowing to store infinitely many patterns $p$.

This implies that for this AED model at any given time
step $t$ all spins are uncorrelated and, hence, the first step dynamics
describes the full time evolution of the network.

For the SED model the architecture is still a local
Cayley-tree  but no longer directed and in  the limit $N \rightarrow
\infty$ the probability that the number of connections
 giving information to the
the  site $i$, is still a Poisson distribution with mean
$C$. However, at time $t$ the spins are no longer uncorrelated
causing a feedback from $t \geq 2$ onwards \cite{WS}, \cite{PZSD}.

\subsubsection{First time step}
In order to solve the dynamics we start with a discussion of the first
time step dynamics, the form of which is independent of the architecture.
So consider a FC network. Suppose that the initial
configuration of the network
$\{\sigma_i(0)\}$ is a collection of i.i.d.r.v.\ with mean
$\E[\sigma_i(0)]=0$, variance $\Var[\sigma_i(0)]=a_0$, and correlated with
only one stored pattern, say the first one $\{\xi^1_i\}$:
\begin{equation}
        \label{eq:init1}
        \E[\xi_i^\mu\sigma_j(0)]=\delta_{i,j}\delta_{\mu,1}m^1_0 A
                \quad m^1_0>0 \, .
\end{equation}
This pattern is said to be condensed.  By the law of large numbers (LLN)
one  gets for the main overlap and the activity at $t=0$
\begin{eqnarray}
        m^1(0)&\equiv&\lim_{N \rightarrow \infty} m^1_N(0)
                \ustr{Pr}{=}\frac1A \E[\xi^1_i \sigma_i(0)]
                = m^1_0
                \label{eq:mo}       \\
        a(0)&\equiv&\lim_{N \rightarrow \infty} a_N (0)
                \ustr{Pr}{=} \E[\sigma_i^2(0)]=a_0
                \label{eq:a0}
\end{eqnarray}
where the convergence is in probability (e.g., \cite{SH}). In order to
obtain the configuration at $t=1$ we first have to calculate the local
field (\ref{eq:h}) at $t=0$. To do this we employ the signal-to-noise
ratio analysis (see, e.g.,\cite{PZFC1}, \cite{BSVZ}). Recalling the 
learning
rule (\ref{eq:JFC}) we separate the part containing the condensed
pattern, i.e., the signal, from the rest, i.e., the noise to arrive at
\begin{equation}
        h_i(\bsigma_{N}(0))
        =
        \xi_i^1 \frac{1}{N A} \sum_{j \neq i}
                                     \xi_j^1 \sigma_j(0)
         +
        \sqrt{\alpha}
        \frac{1}{\sqrt{pA}}
        \sum_{\mu \neq 1 }
                \xi_i^\mu
                \frac{1}{\sqrt{NA}}
                  \sum_{j \neq i}
                       \xi_j^\mu \sigma_j(0) \nonumber\\
\label{eq:F13}
\end{equation}
where $\alpha = p/N$. The properties of the initial configurations
(\ref{eq:init1})-(\ref{eq:a0}) assure us that the summation in the first
term on the r.h.s of (\ref{eq:F13}) converges in the limit
 $N \rightarrow \infty$ to
\begin{equation}
        \lim_{N \rightarrow \infty}
                 \frac{1}{N A} \sum_{j \neq i}
                                             \xi_j^1 \sigma_j(0)
        \stackrel{{Pr}}{=}  m^1(0).
\label{eq:F14}
\end{equation}
The first term $\xi_i^1m^1(0)$ is independent of the second term on the
 r.h.s
of (\ref{eq:F13}). This second term contains the influence of the
non-condensed patterns causing the intrinsic noise in the dynamics of
the main overlap. In view of this we define the residual overlap
\begin{equation}
    r^\mu(t) \equiv \lim_{N \rightarrow \infty} r_{N}^\mu(t)
        =\lim_{N \rightarrow \infty}
                \frac{1}{A\sqrt{N}}\sum_{j}
                \xi_j^\mu\sigma_j(t)
                \quad \mu \neq 1    \,.
        \label{eq:rdef}
\end{equation}
Applying the CLT to this second term in (\ref{eq:F13}) we find
\begin{eqnarray}
   \hspace*{-0.3cm}
     \lim_{N \rightarrow \infty} \sqrt{\frac{\alpha}{p}}
           \sum_{\mu \neq 1}
               \xi_i^\mu r_{N \setminus i}^\mu(0)
        &=& \lim_{N \rightarrow \infty} \sqrt{\alpha}
                \frac{1}{\sqrt{p}}
                \sum_{\mu \neq 1}
                        \xi_i^\mu
                        \frac{1}{A\sqrt{N}}
                        \sum_{j \neq i}
                                \xi_j^\mu \sigma_j(0) \\
           &\stackrel{{\cal D}}{=}&
                 \sqrt{\alpha}~{\cal N}(0,AD(0))
\label{eq:F15}
\end{eqnarray}
where the quantity ${\cal N}(0,V)$ represents a Gaussian random variable
with mean $0$ and variance $V$ and where $D(0)=\Var[r^\mu(0)]=a(0)$.
Thus we see that in fact the variance of this residual overlap, i.e.,
$D(t)$ is the relevant quantity characterising the intrinsic noise.

In conclusion, in the limit $N \rightarrow \infty$ the local field is the
sum of two independent random variables, i.e.
\begin{equation}
        h_i(0)
        \equiv
        \lim_{N \rightarrow \infty} h_{N,i}(0)
        \stackrel{{\cal D}}{=}
        \xi_i^1 m^1(0) + \sqrt{\alpha} {\cal N}(0,a(0)).
\label{eq:F16}
\end{equation}
At this point we note that the structure  (\ref{eq:F16}) of
the distribution of the local field at time zero -- signal plus Gaussian
noise -- is typical for all architectures discussed here because the
correlations caused by the dynamics only appear for $t \geq 1$. Some
technical details are different for the various architectures. The first
change in details that has to be made is an adaptation of the sum over
the  sites
$j$ to all $i$ for the LF architecture and to 
the part of the tree connected to neuron $i$ which has mean $C$, in the 
ED architectures. The second change is that for
the diluted architectures an additional limit $C \rightarrow \infty$
has to be taken besides the $N \rightarrow \infty$ limit. So in the
thermodynamic limit $C, N \rightarrow \infty$ all
averages will have to be taken over the treelike structure, viz.
$\frac{1}{N}\sum_{i } \rightarrow \frac{1}{C} \sum_{i \in tree}$.
Furthermore $\alpha =p/N$ has to be replaced by $\alpha =p/C$.

\subsubsection{Recursive dynamical scheme}

The key question is then how these quantities evolve in time under the
parallel dynamics specified before.
For a general time step we find from  the
LLN in the limit $C,N \rightarrow \infty$ for the main overlap
and the activity (\ref{eq:mdef})
\begin{eqnarray}
        m^1(t+1) \ustr{Pr}{=} \frac{1}{A} \langle\!\langle
                 \xi_i^1\langle\sigma_i(t+1)\rangle_{\beta}
                                  \rangle\!\rangle , \quad
        a(t+1)   \ustr{Pr}{=} \langle\!\langle
              \langle\sigma_i(t+1)\rangle_{\beta}^2
                         \rangle\!\rangle
          \label{eq:aT}
\end{eqnarray}
with the thermal average defined as
\begin{equation}
        \langle f(\sigma_i(t+1)) \rangle_{\beta}
        =
        \frac
        {\sum_{\sigma \in {\cal S}}
        f(\sigma)
        \exp[\frac{1}{2} \beta\,\sigma(h_i(t) - b\sigma)]}
        {\sum_{\sigma \in {\cal S}}
        \exp[\frac{1}{2} \beta\,\sigma(h_i(t) - b\sigma)]}
        \label{eq:thermal}
\end{equation}
where $h_i(t) \equiv \lim_{N \rightarrow \infty} h_{N,i}(t)$.
In the above $\langle\!\langle \cdot \rangle\!\rangle$
denotes the average both over the distribution of the embedded patterns
$\{\xi_i^\mu\}$ and the  initial configurations $\{\sigma_i(0)\}$. The
average over the latter is hidden in an average over the
local field through the updating rule (\ref{eq:gain}).
In the sequel we focus on zero temperature. Then, using
eq.~(\ref{eq:gain}) these formula reduce to
\begin{eqnarray}
        m^1(t+1) \ustr{Pr}{=} \frac{1}{A} \langle\!\langle
                 \xi_i^1\mbox{g}_b(h_i(t)) \rangle\!\rangle , \quad
        a(t+1)   \ustr{Pr}{=} \langle\!\langle \mbox{g}_b^2(h_i(t))
                         \rangle\!\rangle \, .
          \label{eq:a}
\end{eqnarray}

As seen already in the first time step, we have to study carefully the
influence of the non-condensed patterns causing the intrinsic noise in
the dynamics of the main overlap.
The method used to obtain these order parameters is then to calculate
the  distribution of the local field as a function of time.
In order to determine the structure of the local field we
have to concentrate on the evolution of the residual overlap. The
details of  this calculation are very technical and depend on the
precise correlations in the system and hence on the architecture of the
network \cite{BJSF}, \cite{BVZ}, \cite{BSVZ}, \cite{BSV}, \cite{BJSsymdil}.
Here we give a  discussion of the results obtained.

In general, the distribution of the local field at time $t+1$ is given by
\begin{equation}
        h_i(t+1)=\xi_i^1m^1(t+1) + {\cal N}(0,\alpha a(t+1))
           + \chi(t) [F(h_i(t)-\xi_i^1m^1(t))+B\alpha\sigma_i(t)]
              \label{eq:hrec}
\end{equation}
where $F$ and $B$ are binary coefficients given below, which depend on the
specific architecture. {}From this it is clear that the local field at time
$t$ consists out of a discrete part and a normally distributed part, viz.
\begin{equation}
        h_i(t)=M_i(t) + {\cal N}(0, V(t))
\end{equation}
where $M_i(t)$ satisfies the recursion relation
\begin{equation}
        M_i(t+1)=\chi(t) [F(M_i(t)-\xi_i^1m^1(t))+B\alpha\sigma_i(t)]
                         + \xi_i^1m^1(t+1)
     \label{eq:Mrec}
\end{equation}
and where $V(t)=\alpha A D(t)$ with $D(t)$ itself given by the
recursion relation
\begin{equation}
        \label{eq:Drec}
        D(t+1)=\frac{a(t+1)}{A}+L \, \chi^2(t)D(t)+
                2 B_2 \, \chi(t) {Cov}[\tilde r^\mu(t),r^\mu(t)]
 \label{eq:f4}
\end{equation}
where $L$ and $B_2$ are again coefficients specified below. The
quantity $\chi (t)$ reads
\begin{equation}
        \chi(t) = \sum_{k=1}^{Q-1} f_{\hat h_i^\mu (t)}(b(s_{k+1}+s_k))
                   (s_{k+1}-s_k)
            \label{eq:chi}
\end{equation}
where $f_{\hat h_i^\mu (t)}$ is the probability density of
 $ \hat h_i^\mu (t) = \lim _{N \to \infty} \hat h_{N,i}^\mu(t)$ with
\begin{equation}
        \hat h_{N,i}^\mu(t)=h_{N,i}(t)-
                \frac{1}{\sqrt{N}}\xi_i^\mu r_N^\mu(t) \, .
        \label{eq:f3}
\end{equation}
Furthermore, $\tilde r^\mu(t)$ is defined as
\begin{equation}
        \tilde r^\mu(t) \equiv \lim_{N \rightarrow \infty}
           \frac1{A\sqrt{N}}\sum_{i} \xi_i^\mu
                \mbox{g}_b(\hat h_{N, i}^\mu(t)) \, .
           \label{eq:w}
\end{equation}
At this point we remark that we made the technical assumption that 
$\sigma_i(t)$ and $\hat h_{N, i}^\mu(t)$ are only weakly correlated in
the limit $N \to \infty$ such that $\tilde r^\mu(t)$ converges to a
normal distribution. 
Finally, as can be read off {}from eq.~(\ref{eq:Mrec}) the quantity
$M_i(t)$  consists out of the signal term and a discrete noise
term, viz.
\begin{equation}
        M_i(t)=\xi _i^1 m^1(t) + B_1\alpha \chi(t-1)\sigma _i(t-1)
        + B_2\sum_{t'=0}^{t-2} \alpha
         \left[\prod_{s=t'}^{t-1} \chi(s)\right] \, \sigma _i(t')  \,.
         \label{eq:MM}
\end{equation}
Since different architectures contain different correlations not all
terms in these final equations are present. In particular we have
for the coefficients $F,B,L,B_1$ and $B_2$ introduced above
\begin{equation}
\begin{array}{l|ccccc}
&F&B&L&B_1&B_2\\
\hline
FC& 1&1&1&1&1\\
SED&0&1&0&1&0\\
LF& 1&0&1&0&0\\
AED&0&0&0&0&0
\end{array}
\end{equation}
with $B$ indicating the feedback caused by the symmetry in the
architectures and $L$ the common ancestors contribution.

At this point we remark that in the so-called theory of statistical
neurodynamics \cite{Nishimori}, \cite{AM}, \cite{Okada1996} one starts 
from a different 
approximate local field by leaving out any discrete noise (the term in
$\sigma_i(t)$). As a consequence the covariance in the recursion relation
for $D(t)$ can be written down more explicitly since only Gaussian noise
is involved. For more details we refer to \cite{Nishimori}, 
 \cite{BBVp}, \cite{jongen}.

We still have to determine the probability density of
$f_{\hat h_i^\mu(t)}$ in eq.~(\ref{eq:chi}), which in the thermodynamic
limit equals the probability density of $f_{h_i(t)}$. This can
be done by looking at the form of $M_i(t)$ given by eq.~(\ref{eq:MM}).
The evolution equation tells us that $\sigma _i(t')$ can be replaced by
$g_b(h_i(t'-1))$ such that the second  and third terms of $M_i(t)$ are
the sums of stepfunctions of correlated variables. These are also
correlated through the dynamics with the normally distributed
part of $h_i(t)$. Therefore the local field can be considered as a
transformation of a set of correlated normally distributed variables
$x_s,\, s=0,\ldots,t-2,t$, which we choose to normalize. Defining the
correlation matrix $W = \left(\rho(s,s')\equiv \E[x_s x_{s'}] \right)$
we  arrive at the following expression for the probability density of
the  local field at time~$t$
\begin{eqnarray}
     f_{h_i(t)}(y)&=&\int\prod_{s=0}^{t-2} dx_s dx_t ~
         \delta \left(y - M_i(t)-\sqrt{\alpha A D(t)}\,x_t\right)
             \nonumber\\
             &\times& \frac{1}{\sqrt{\mbox{det}(2\pi W)}}
            ~\mbox{exp}\left(-\frac{1}{2}{\bf x} W^{-1}
            {\bf x}^T\right)
            \label{eq:fhdisfc}
\end{eqnarray}
with ${\bf x}=(x_0,\ldots x_{t-2},x_t)$. For the symmetrically diluted
case this expression simplifies to
\begin{eqnarray}
     f_{h_i(t)}(y)&=&\int\prod_{s=0}^{[t/2]} dx_{t-2s} ~
             \delta \left(y -\xi^1_i m^1(t)- \alpha \chi(t) \sigma_i(t)
              -\sqrt{\alpha a(t)}\,x_t\right) \nonumber\\
             &\times& \frac{1}{\sqrt{\mbox{det}(2\pi W)}}
            ~\mbox{exp}\left(-\frac{1}{2}{\bf x} W^{-1}
            {\bf x}^T \right)
            \label{eq:fhdisd}
\end{eqnarray}
with ${\bf x}=(\{x_s\})=(x_{t-2[t/2]},\ldots x_{t-2},x_t)$. The brackets
$[t/2]$ denote the integer part of $t/2$.

So the local field at time $t$ consists out of a
signal term, a discrete noise part and a normally distributed noise part.
Furthermore, the discrete noise and the normally distributed noise are
correlated and this prohibits us to derive a closed expression for the
overlap and activity.

Together with the eqs.~(\ref{eq:a}) for $m^1(t+1)$ and $a(t+1)$ the
results above form a recursive scheme in order to obtain the
order parameters of the system. The practical difficulty which remains is
the explicit calculation of the correlations in the network at different
time steps as present in eq.~(\ref{eq:f4}).

For AED and LF architectures this scheme leads to an explicit form for
the recursion relations for the order parameters
\begin{eqnarray}
  m^\mu(t+1)
   =
   \frac{\delta_{\mu,1}}{A} \left\langle\!\left\langle\xi^1(t+1)
   \int {\cal D}z\,\mbox{g}_b(\xi^1(t+1)m^1(t)+\sqrt{\alpha AD(t)}\,z)
   \right\rangle\!\right\rangle\label{eq:rm} \\
 a(t+1)
   =
   \left\langle\!\left\langle \int {\cal D}z\,\mbox{g}_b^2(\xi^1(t+1)m^1(t)
    +\sqrt{\alpha AD(t)}\,z) \right\rangle\!\right\rangle
    \label{eq:ra} \\
  D(t+1)
   =
   \frac{a(t+1)}{A}+\frac{L}{\alpha A} \left\langle\!\left\langle
         \int {\cal D}z\,z\mbox{g}_b(\xi^1(t+1)m^1(t)
        +\sqrt{\alpha AD(t)}\,z)\right\rangle\!\right\rangle^2
    \nonumber \\ \label{eq:rD}
\end{eqnarray}
with ${\cal D}z = dz (2 \pi)^{-1/2} \exp(-z^2/2)$.
For the AED architecture $(L=0)$ the second term on the r.h.s. of 
(\ref{eq:rD})
coming from the correlations caused by the common ancestors is absent.
For the LF architecture we remark that this explicit solution requires an
independent choice of the representations of the patterns at different
layers. 

At finite temperatures analogous recursion relations for the AED
and LF networks can be
derived \cite{BSVZ}, \cite{BSV} by introducing auxiliary thermal fields
\cite{PZparis} in order to express the stochastic dynamics within the
gain function formulation of the deterministic dynamics. These recursion
relations can be solved numerically and the stationary limit can be
discussed (see Section 2.2.4). Furthermore,
damage spreading \cite{DGZ}, \cite{D}, \cite{DM}, i.e., the evolution of 
two network
configurations which are initially close in Hamming distance can be
studied \cite{BSVZ}, \cite{BSV}. Finally, a complete self-control 
mechanism can
be built in the dynamics of these systems by introducing a
time-dependent threshold in the gain function improving, e.g., the 
storage capacity, the basins of
attraction of the embedded patterns and the mutual information content
 \cite{Okada1996}--\cite{BD00}, \cite{BDA}

For the symmetric networks explicit examples of the dynamical scheme 
above and a comparison with numerical simulations have been presented
in \cite{BJSF} for the FC and in \cite{BJSsymdil} for the SED
model with equidistant states and a uniform distribution of the patterns.
By using the recursion relations the first few time steps are 
written out explicitly and studied numerically. These results are 
compared with the literature
\cite{K}, \cite{AM}, \cite{PZFC1}, \cite{PZFC2}, \cite{WS}, \cite{PZSD},
\cite{Okada1996}, \cite{GDM}--\cite{GSZ}
 where the feedback correlations 
for $t \geq 2$ are neglected or approximated in different ways.
In the whole retrieval region of these symmetric networks it is found 
that the first
four or five time steps calculated by the scheme presented above give 
already a clear picture of the time evolution. Explicit results 
depend of course on the specific values of the model parameters, e.g., 
the storage capacity $\alpha$, the initial
overlap $m_0$ with the embedded pattern, the initial neural activity
$a_0$, the value of the gain parameter $b$. Furthermore, numerical 
simulations
provide good support for this scheme, but very recently we have discovered
some small deviations, especially close to the border of retrieval which
can not be entirely due to finite size effects. This has been completely 
understood recently by carefully studying the long time correlations
and the details are being worked out (see \cite{BBVp}, 
\cite{V} and Section 3.2).

\subsubsection{Stationary limit: thermodynamic and retrieval properties}
 \label{sec:fixp}

Equilibrium results for the AED and LF $Q$-Ising models are obtained
immediately by straightforwardly leaving out the time dependence in
(\ref{eq:rm})-(\ref{eq:rD}) (cfr. \cite{BSVZ},\cite{BSV}), since the
evolution  equations for the local field and the order parameters
do not change their form as time progresses.  This still allows small 
fluctuations
in the configurations $\{\sigma_i \}$.  The difference between the
fixed-point equations for these two architectures is that
for the AED model the variance of the residual noise, $D(t)$, is
simply proportional to the activity of the neurons at time $t$ while
for  the LF model a recursion is needed.

A lot of detailed results are available on capacity-gain parameter
and temperature capacity diagrams obtained by numerically solving these
equations. In general, it is necessary to distinguish three different
types of solutions. The zero solution, $Z$, is determined by $m=0$  
as well as $a=0$. A sustained activity solution, $S$, is defined by
$m=0$  but $a \neq 0$. Finally, there are solutions with both $m \neq0$
and $a \neq 0$.  Nonattracting solutions of the last type are denoted by
$NR$ (for nonretrieval), attracting ones by $R$ (for retrieval).
As a typical illustration we show fig.~1. 

\begin{figure}
\includegraphics[width=.9\textwidth,clip=]{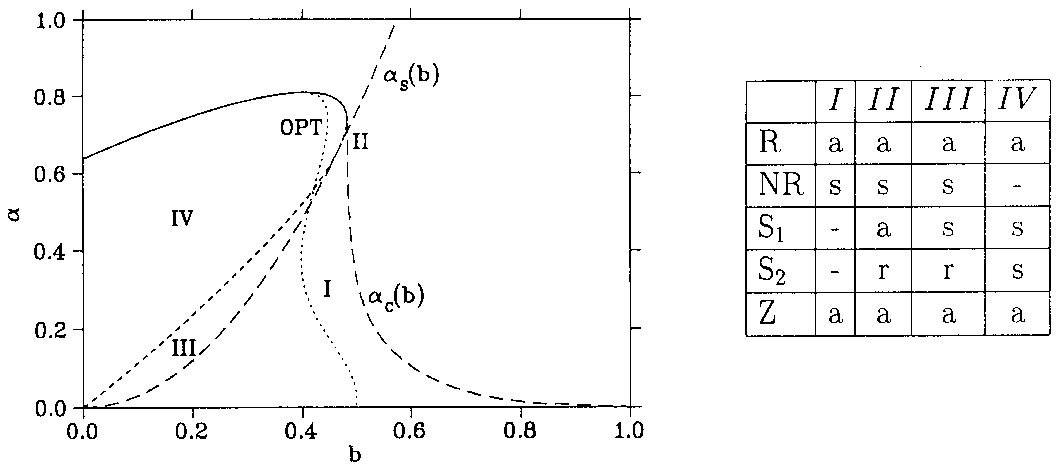}
\includegraphics[width=.75\textwidth,clip=]{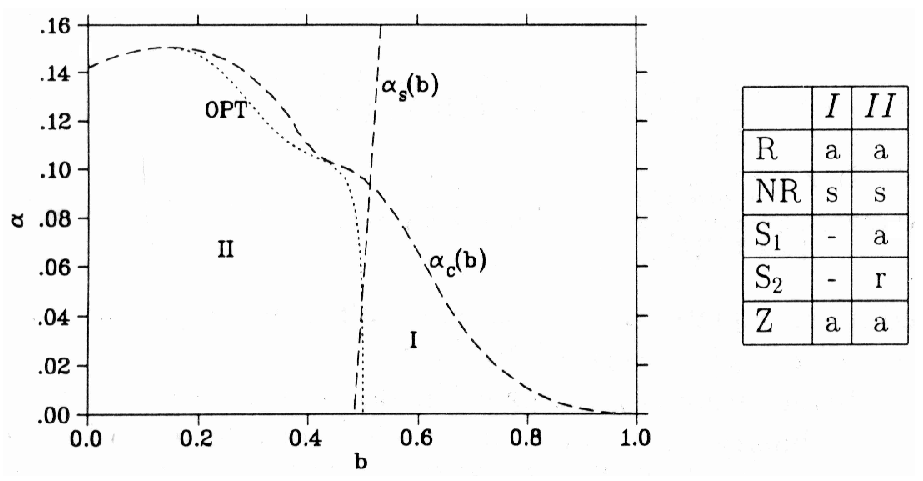}
\caption{ \small The ($\alpha-b$) diagram for the $Q=3$ AED (top) and LF
 (bottom)
network with uniform patterns at $T=0$. The curve $\alpha_c(b)$ denotes
the boundary of the retrieval region. The curve $\alpha_S(b)$ is the
lower bound for the existence of the sustained activity states. The full
line denotes a second-order transition, the dashed line a first order
one. The line $OPT$ is the line of best retrieval quality. The structure
of the retrieval dynamics is explained: a denotes an attractor, s a
saddle-point, r a repellor.}
\end{figure}

For the AED architecture it is important to observe that, for zero
temperature, in the retrieval regime, $R$ is never
the only attractor in the $(m,a)$ plane. Its basin of attraction is
always limited by at least one attractor on the axis $m=0$. 
In contrast, in the case of analog neurons (piecewise linear networks) 
the retrieval solution is an attractor for the whole $(m,a)$ plane.
Furthermore,
at any fixed $\alpha$, a value of $b$ can be determined where the
Hamming  distance of $R$ is minimal. The line of these optimal $b$ is
indicated by $OPT$. It is close to $1/2$ for $T=0$ and shifts completely
to $b=0$ with increasing temperature.   
Finally, for a finite $Q$ network two arbitrarily close configurations
always repel each other even in the retrieval regime. For analog
networks there exists a transition line in the capacity-gain plane below
which no such ``chaotic'' behavour occurs.

For the LF architecture at zero temperature, in contrast to the AED case,
 the
retrieval state is always accompanied by an attractor which has zero
overlap with the embedded pattern. In all cases under consideration the 
retrieval state disappears discontinuously as the storage capacity
$\alpha$ increases. Finally, a type of chaoticity in the network 
trajectories is always present for arbitrary
finite $Q$. However, in the case of a piecewise linear gain function
there exists a dynamical transition towards chaos in the
$(\alpha,b)$-plane.
The $(\alpha,b)$-region where chaos does occur is relatively smaller than
in the corresponding AED networks. For further results and especially
for results at finite temperature,  we refer to the
literature mentioned before.

Next, for the SED and FC architectures  the evolution equations for the
order parameters do change their form  as time progresses by the explicit
 appearance of the
$\{\sigma_i(t')\}, t'=1,\ldots,t $ term.  Hence we can not use the
simple procedure above to obtain the fixed-point equations. Instead we
derive the equilibrium results of our dynamical scheme by requiring
through the recursion relations (\ref{eq:hrec}) that the distribution
of the local field becomes time-independent. This is clearly an 
approximation
because fluctuations in the network configuration are no longer allowed.
In fact, it means that out of the discrete part of
this distribution, i.e., $M_i(t)$ (recall (\ref{eq:MM})), only the
$\sigma _i(t-1)$ term is kept besides, of course, the signal term.
This procedure implies that the main overlap and activity in the
fixed-point are found {}from the definitions (\ref{eq:mdef})
and not {}from leaving out the time dependence in the recursion relation
(\ref{eq:a}).

We start by eliminating the time-dependence in the evolution equations for
the local field (\ref{eq:hrec}). This leads to
\begin{equation}
        \label{eq:hfix}
        h_i=\xi_i^1m^1 + [{\bar \chi}^{ar}]^{-1} {\cal N}(0,\alpha a)
                +[{\bar \chi}^{ar}]^{-1}\alpha \chi \sigma_i
\end{equation}
with ${\bar \chi}^{ar} \equiv 1-F\chi$ being $1$ for the SED and $1-\chi$ 
for the FC model and
$h_i \equiv \lim_{t \rightarrow \infty} h_i(t)$.
This expression consists out of two parts: a normally distributed part
$\tilde h_i = {\cal N}(\xi_i^1m^1,\alpha a / [{\bar \chi}^{ar}]^2)$
and some discrete noise part. The
discrete noise comes from the correlations of the $\{\sigma_i(t)\}$ at
different time steps (here only the preceding time step is considered)
and is inherent in the SED and FC dynamics. 
Employing this expression in the updating rule (\ref{eq:gain}) one finds
\begin{equation}
        \label{eq:sfp}
        \sigma_i=\mbox{g}_b(\tilde h_i+
                [{\bar \chi}^{ar}]^{-1} \alpha \chi \sigma_i)\,.
\end{equation}
This is a self-consistent equation in $\sigma_i$ which
in general admits more than one solution. These types of equation have been
solved in the literature in the context of thermodynamics using a
geometric Maxwell construction \cite{SFb}, \cite{SFa}.
We remark that for analog networks
the geometric Maxwell construction  is not necessary: the
fixed-point equation (\ref{eq:sfp}) has only one solution
 \cite{jongen}.

This approach leads to a unique solution
\begin{equation}
        \sigma_i=\mbox{g}_{\tilde b}(\tilde h_i), \quad
        {\tilde b} = b - [{2\bar \chi}^{ar}]^{-1}\alpha \chi
         \label{btilde}   \,.
\end{equation}
We remark that plugging this result into the local field (\ref{eq:hfix})
tells us that the probability distribution of the local field contains
$(Q-1)$ gaps. This gap structure also depends on the architecture and the 
most important findings are that dilution changes the regions of
existence of these gaps but not their width. Moreover, the gaps become
typically much bigger when crossing the border of retrieval 
\cite{BS}--\cite{CS}.

Using the definition of the main overlap and activity
(\ref{eq:mdef}) in the limit $N \rightarrow \infty$
for the FC model and limit $C,N \rightarrow \infty$ for the SED model,
one finds in the fixed point
\begin{eqnarray}
        \label{eq:m1fix}
        m^1 =\left\langle\!\left\langle\xi^1\int {\cal D}
        z ~  \mbox{g}_{\tilde b}
                \left( \xi^1m^1 + \sqrt{\alpha A D}\,z
                \right)\right\rangle\!\right\rangle \,
         \\
        \label{eq:afix}
        a   =\left\langle\!\left\langle\int {\cal D}
        z ~  \mbox{g}_{\tilde b}^2
                \left( \xi^1m^1 + \sqrt{\alpha A D}\,z
                \right)\right\rangle\!\right\rangle
          \,.\end{eqnarray}
From (\ref{eq:Drec}) and (\ref{eq:chi}) one furthermore sees that
\begin{equation}
        \label{eq:Dfix}
        D = [{\bar \chi}^{ar}]^{-2} a/A
\end{equation}
with
\begin{equation}
        \label{eq:chifix}
        \chi=\frac1{\sqrt{\alpha A D}}
                \left\langle\!\left\langle\int {\cal D}
                z ~  z \, \mbox{g}_{\tilde b}
                        \left( \xi^1m^1 + \sqrt{\alpha A D}\,z
                        \right)\right\rangle\!\right\rangle \,.
\end{equation}
These resulting equations (\ref{eq:m1fix})-(\ref{eq:Dfix}) obtained
through {\it parallel} dynamics turn out to
be  the same as the fixed-point equations derived from a 
replica-symmetric mean-field theory treatment  discussed next.

For symmetric networks (FC and SED) we consider the long time behaviour 
governed by the Hamiltonian
\begin{equation}
         H^{ar} = -\frac12\sum_{i\neq j} J_{ij}^{ar}\sigma_i\sigma_j
                         +b\sum_i \sigma_i^2\,.
        \label{eq:s1}
\end{equation} 
with $J_{i,j}^{ar}$ given by (\ref{eq:JFC}) for the FC and by
 (\ref{eq:JED}) for the SED model. The neurons
are updated {\it asynchronously} according to the transition probability
(\ref{eq:trans})-(\ref{eq:energy}). In order to calculate the free energy 
we use the standard replica method \cite{AGS}, \cite{MPV}. We remark
that  for the SED architecture, we employ the replica method as
applied to dilute spin-glass models \cite{WS2}--\cite{bcs76}. Starting
from the replicated partition function averaged 
over the connectivity  and the non-condensed patterns and
assuming replica symmetry, we arrive at the free energy $f(\beta)$ which 
can be written down for a variable dilution $c=C/N$ with $c$ between 
$0$ (SED) and $1$ (FC)
\begin{eqnarray}
&&  f(\beta) =
   \frac{A}{2}\sum_{\mu=1}^s \left( m_{\mu} \right)^2 
        + \frac{ \alpha c}{2 \beta} \left[\ln(1-\chi) 
	              + \frac{\chi}{1-\chi} 
        + \frac{ q \beta \chi}{(1-\chi)^2} \right] 
             \nonumber     \\
 &&-
  \frac{1}{\beta}
  \left\langle\kern-0.3em\left\langle 
      \int {\cal D} z 
      \ln {\mbox Tr}_{ \{ \sigma \}}
         \exp\left[\beta \sigma
	    \left(\sum_{\mu} m_{\mu} \xi^{\mu}+ z \sqrt{\alpha rc}  
                - {\tilde b}\sigma \right)
             \right]
          \right\rangle\kern-0.3em\right\rangle_{\{\xi\}} 
	\label{freenew}
\end{eqnarray}
with $s$ the number of condensed patterns which we  take to be $1$ as 
before,
\begin{equation}
  {\tilde b}= b- \frac{\alpha \chi}{2}
                \left[1+  \frac{c\chi}{1-\chi}\right] \,,
  \quad
  r=q \left[ \frac{1}{(1-\chi)^2} + \frac{1-c}{c}  \right]
\end{equation}
with $q$ the Edwards-Anderson spin-glass order parameter and  
$\chi= \beta\left \langle \! \left \langle \langle \sigma^2 \rangle
- \langle \sigma \rangle^2 \right \rangle \! \right \rangle $ 
the susceptibility (defined before in Section 2.2.3 for zero
temperature) in the stationary limit. We remark that the effective gain 
parameter $\tilde b$ can be negative,
implying that the input-output function reduces to that of $2$-Ising-type
neurons, i.e., $g_{\tilde{b}}(h) = \mbox{sign}(h)$.

The phase structure of the network is determined by the
solution of the fixed-point equations for the order parameters
\begin{eqnarray}
    m_\mu &=& \frac1A \left \langle \! \left \langle
             \int Dz\, \xi^\mu \langle{\sigma(z)} \rangle
             \right \rangle \! \right \rangle
              \label{eq:s6} \\
    q &=& \left \langle \! \left \langle
               \int Dz\, \langle{\sigma(z)}^2\rangle
               \right \rangle \! \right \rangle
               \label{eq:s7} \\
    \chi &=& \frac{1}{\sqrt{\alpha rc}} 
    \left \langle \! \left \langle
        \int Dz\, z\,\langle{\sigma(z)} \rangle 
	   \right \rangle \! \right \rangle 
                \label{eq:s8}
\end{eqnarray}
which maximize $-\beta f (\beta)$. Here
\begin{equation}
      \langle{\sigma(z)}\rangle = \frac{{\mbox Tr}_{\sigma} \sigma \exp[\,
           \beta\sigma\,(\sum_\mu m_\mu \xi^\mu+ 
	   z \sqrt{\alpha rc} -\tilde{b}\sigma)]}
      { {\mbox Tr}_s \exp[\, \beta s\,(\sum_\mu
      m_\mu \xi^\mu + z \sqrt{\alpha rc}-\tilde{b} s)]}\,.
           \label{eq:s9}
\end{equation}
Explicit expressions for these fixed-point equations  for $Q=3,4$ and 
$Q=\infty$ can be found in \cite{R}, \cite{BRS}, \cite{BCS00} for the
FC and SED model and for $Q=3$ in \cite{TE01} for variable dilution. 

A lot of detailed results are available on the corresponding phase
diagrams. Some typical results are shown in fig.~2.
\begin{figure}[t]
\includegraphics[width=.49\textwidth,clip=]{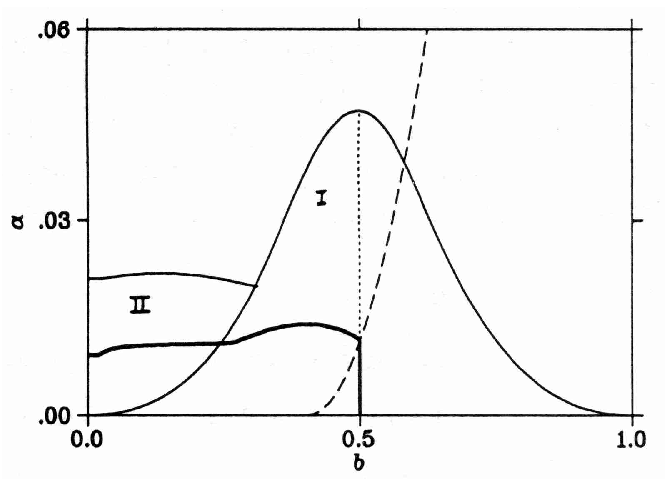}
\includegraphics[width=.49\textwidth,clip=]{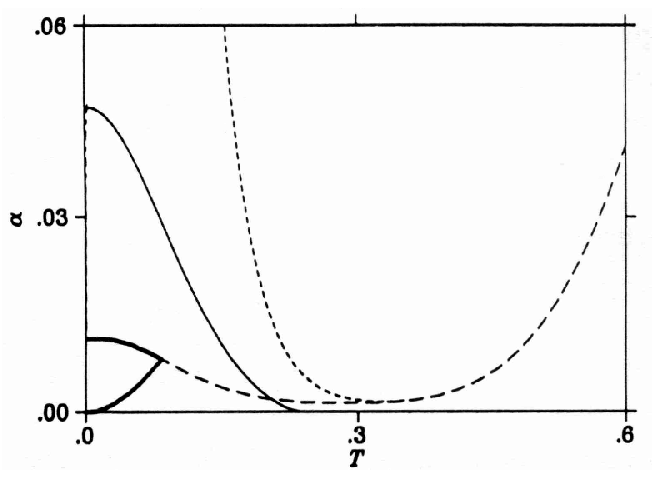}
   \caption{\small {The $(\alpha -b$), $T=0$ (left) and $(\alpha -T)$,
    $b=\frac12$ (right) phase diagram for the FC $Q=3$-Ising model with
     uniform patterns. The (thin) full curve represents the boundary 
     of the retrieval region, the
      thick full curve the thermodynamic transition of the retrieval 
      state, the long-dashed curve the spin-glass transition, and the 
      dotted curve the optimal gain parameter. The I and II indicate 
      two retrieval regions: in region I $r \approx O(1)$, while in 
      region II $r \approx O(10)$. The chain curve (very close to the
      $\alpha$-axis on the right)is the AT-line. The short dashed curve 
      indicates
      the border above which no paramagnetic states exists. 
       }}
\end{figure} 
\begin{figure}[t]
\begin{center}
\includegraphics[width=.5\textwidth,clip=]{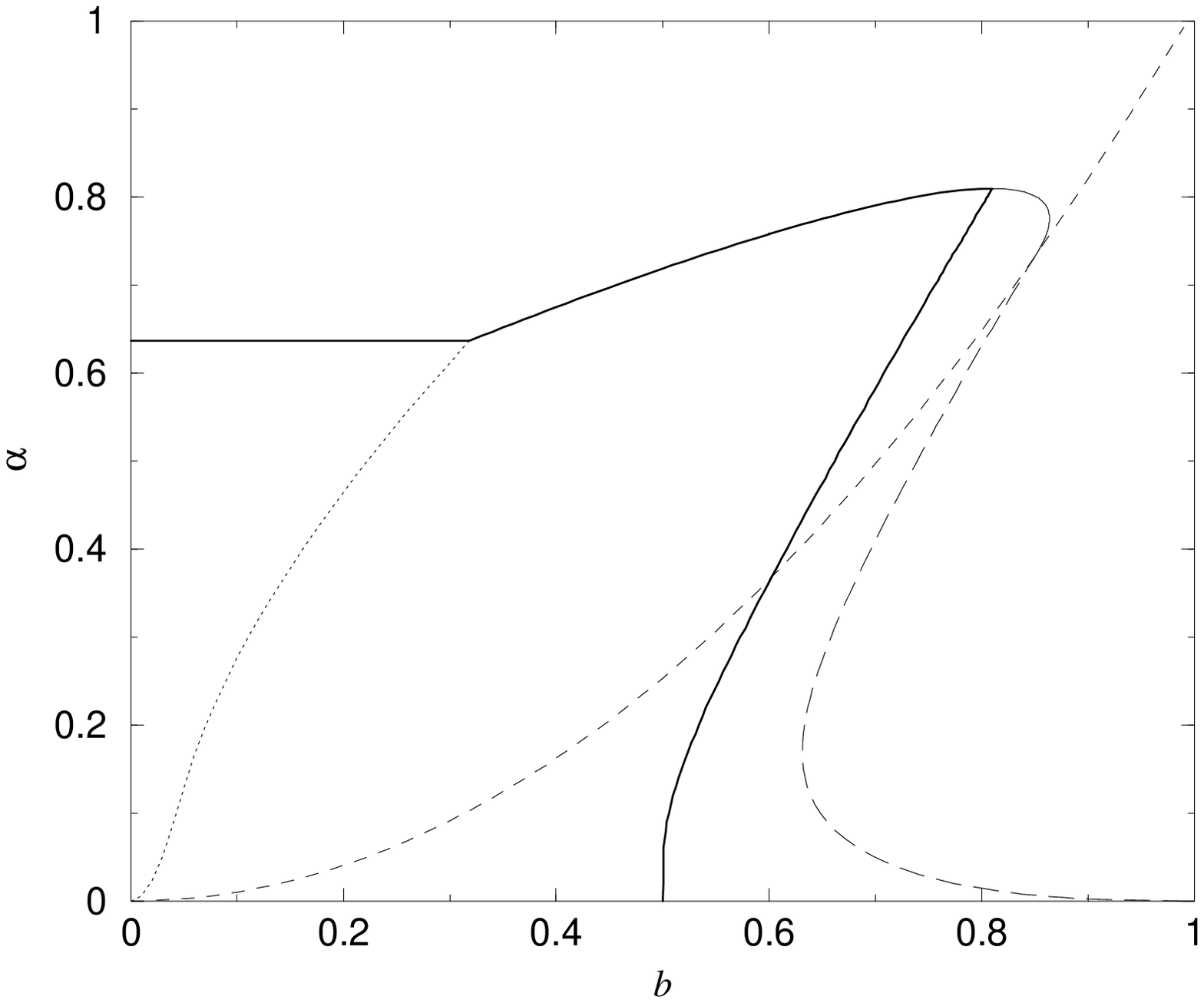}
\end{center}
   \caption{\small The $(\alpha -b)$ phase diagram for the $Q=3$ SED model 
   with uniform patterns at $T=o$. 
      The (thin) full and long-dashed curves denote the boundary of the
      retrieval region corresponding  respectively to a continuous and
      discontinuous appearance of the solution. The dotted curve separates
      the $2$-Ising-like retrieval region. The short-dashed curve 
      indicates the discontinuous spin-glass transition.  
      The thick full curve represents the thermodynamic transition for the
      retrieval state.}
\end{figure} 
In general, we can distinguish a retrieval phase ($m \neq 0$), a
spin-glass phase ($m=0, q>0$) and a paramagnetic phase ($m=0, q=0$).
For the FC architecture at zero temperature the results are extremely 
dependent on the pattern activity. In the case of uniformly distributed 
patterns ($A=2/3$) we see that different retrieval regions show up for 
small $b$ (the retrieval region II does not appear for $A<1/3$) and 
the capacity is reduced by a factor compared with the capacity for the
AED and LF architectures. The line of optimal Hamming distance is given
exactly by $b=1/2$ (fig~2 left); in the AED model we recall that it is 
located for the whole retrieval region in the interval $b \in [0.4,0.5]$ 
(see fig.~1 top) while in the LF model it bends to smaller $b$ for 
growing $a$ (see fig.~1 bottom). We remark that for $Q= \infty$ the 
diagram for the FC and LF models are very similar in shape but the 
capacity is reduced roughly by a factor of $10$ in the former.
For non-zero temperatures the situation is complicated
and depends very much on the value of $b$. For $b$ close to and greater
than the optimal $b=1/2$ the phase diagram is completely different from
that of the Hopfield model in the sense that the paramagnetic phase
exists between the retrieval and the spin-glass phase (see fig.~2 right). 
We remark that for $Q=\infty$ the diagram is relatively simple again 
and qualitatively resembles that of the Hopfield model.

For the SED architecture there are interesting similarities with the AED
model. In fact, we find that the $(\alpha-b)$ phase diagram in fig.~3 is 
tilted  towards higher $b$-values in comparison with fig.~1 (top) because 
of the presence of  
an extended $2$-Ising-like region. The critical boundary of this region
is independent of $Q$. The $(\alpha-T)$ diagram of the $Q=3$  and
$Q=\infty$ models are qualitatively similar.

For variable dilution (all values of $c \in [0,1]$) one finds some
architecture independent properties for $\alpha \to 0$, e.g., the optimal
value of $b$ being $b=1/2$ for $T=0$. The main dependence of the
behaviour of the network on the connectivity arises for finite $\alpha$.
An interesting property is the suppression of the discontinuous boundary
between the retrieval regions I and II (see fig.~2 left) with decreasing 
connectivity, disappearing completely for $c \approx 0.63$, making the 
optimal performance domain readily accessible to a wide region of network
parameters \cite{TE01}.  
 
Finally, the stability of the replica symmetric retrieval solution against
replica-symmetry breaking can be determined by studying the replicon
eigenvalue \cite{MPV} \cite{AT}, leading to the de Almeida-Thouless (AT)
stability line indicating the temperatures below which the
replica-symmetric approximation is no longer valid. For more details we 
refer to the figures shown and to the literature mentioned before.

\section{BEG neural networks} \label{sec:modbeg}
\subsection{The model}

In Section 2.1 it has been mentioned  that the mutual information
\cite{Sh48}, \cite{Bl90} is the 
most appropriate concept to measure the retrieval quality for
sparsely coded networks. A natural question is then whether one could use
the mutual information in general in a systematic way to determine a priori
an optimal hamiltonian guaranteeing the best retrieval properties
including, e.g., the largest retrieval overlap, loading capacity, basin
of attraction, convergence time,  for an arbitrary scalar valued neuron 
(spin) model. Optimal means especially that although the network might  
start initially
{\it far} from the embedded pattern it is  still able to retrieve it.

This question can be answered positively  \cite{DK00}, \cite{BV02} 
by presenting a general scheme in order to express the mutual information
as a function of the
relevant macroscopic parameters like, e.g., overlap with the embedded
patterns, activity, $\ldots$ and constructing a hamiltonian from it for 
general $Q$-state neural networks.
For $Q=2$, one finds back the  Hopfield model for biased patterns
 ensuring that this hamiltonian is optimal in the sense
described above. For $Q=3$, one obtains a Blume-Emery-Griffiths (BEG) type 
hamiltonian \cite{DK00} named after the BEG spin-glass \cite{ACN00}. 

This BEG-model for a FC architecture can then be descibed as follows.
Consider a neural network consisting of $N$ neurons which can take
values $\sigma_i, i=1,\ldots, N$ from the discrete set
$ \mathcal{S}\equiv \lbrace -1,0,+1 \rbrace $.
The $p=\alpha N$ patterns to be stored in this network are supposed to
be i.i.d.r.v., $\{\xi_i^\mu\} $, $\mu =1,\ldots,p$ 
with a probability distribution
\begin{equation}
p(\xi_{i}^{\mu})=\frac{a}{2}\delta(\xi_{i}^{\mu}-1)+
   \frac{a}{2}\delta(\xi_{i}^{\mu}+1)+(1-a)\delta(\xi_{i}^{\mu})
   \label{begprobability}
\end{equation}
with $a$ the activity of the patterns so that
\begin{equation}
  \lim_{N\rightarrow\infty}\frac{1}{N}\sum_{i}(\xi_{i}^{\mu})^2 = a.
\end{equation}
(We remark that for reasons of convenience the pattern activity in this
Section  is now
indicated with $a$ and not with $A$ as in the $Q$-Ising Section.)

Given the network configuration at time $t$,
${\bsigma}_N(t)\equiv\{\sigma_j(t)\}, j=1,\ldots,N$,
the following dynamics is considered.   
The configuration $\bsigma_N(0)$ is chosen as input. 
All neurons are updated in parallel according to the
rule (\ref{eq:enpot}) at zero temperature  or the transition probability 
(\ref{eq:trans}) at arbitrary temperature.
But, here the energy potential  $\epsilon_i[s|{\bsigma}_N(t)]$ is 
different from (\ref{eq:energy}) and defined by
\begin{equation}
          \epsilon_i(s|{\bsigma}_N(t)) = 
     -sh_i({\bsigma}_N(t))-s^2\theta_i({\bsigma}_N(t))
               \,, \label{eq:begenergy}
\end{equation}
where the following local fields in neuron $i$ carry all the information
\begin{equation}
        \label{eq:begh}
      h_{N,i}(t)=\sum_{j \neq i} J_{ij}\sigma_j(t), \quad
      \theta_{N,i}(t)=\sum_{j\neq i}K_{ij}\sigma_{j}^{2}(t)
\end{equation}
with the obvious shorthand notation for the local fields. The synaptic 
couplings $J_{ij}$ and $K_{ij}$ are of the Hebb-type
\begin{equation}
J_{ij}=\frac{1}{a^{2}N}\sum_{\mu=1}^{p}\xi_{i}^{\mu}\xi_{j}^{\mu},
\qquad 
K_{ij}=\frac{1}{N}\sum_{\mu=1}^{p}\eta_{i}^{\mu}\eta_{j}^{\mu}
\label{beghebb}
\end{equation}
with
\begin{equation}
\eta_{i}^{\mu}=\frac{1}{a(1-a)}((\xi_{i}^{\mu})^{2}-a).
\end{equation} 
The first part is the usual rule in a three-state network (recall
Section 2.1)
that codifies the patterns, while the second part can be considered as
codifying  the
fluctuations of the binary active patterns $(\xi^{{\mu}}_i)^{2}$
about their average. That part is also consistent with the modified
Hebb rule for the Hopfield model with biased patterns. The 
updating rule (\ref{eq:enpot}) is equivalent to using a gain function 
\begin{equation}
        \label{eq:beggain}
        \sigma_i(t+1)  = 
               \mbox{g}(h_{N,i}(t), \theta_{N,i}(t))=
         \mbox{sign}(h_{N,i}(t)) \Theta(|h_{N,i}(t)| + \theta_{N,i}(t))
\end{equation}
with $\Theta$ the Heaviside function.

The order parameters of this system have been obtained starting form the
mutual information as a measure for the retrieval quality of the system 
\cite{DK00}, \cite{BV02}. They are the retrieval overlap, the activity 
overlap, and  the neural activity 
\begin{eqnarray}
        \label{eq:begmdef}
        m_N^\mu(t)&=&\frac{1}{aN}
                \sum_{i}\xi_i^\mu\sigma_i(t),
                \quad  
        n_N^\mu(t)=\frac{1}{aN}\sum_{i}(\xi_i^\mu)^2(\sigma_i(t))^2, 
	      \nonumber   \\
       q_N(t)&=&\frac{1}{N}\sum_{i}(\sigma_{i}(t))^2 \,.
\end{eqnarray}
(We remark that in this Section the neural activity is now denoted by
$q$ instead of $a$.)
Instead of using the activity overlap $n_N^\mu(t)$ itself it is more 
convenient to employ the modified activity overlap
\begin{equation}
l_N^\mu(t)=\frac{1}{1-a}(n_N^\mu(t) - q_N(t))
  =\frac{1}{N}\sum_{i}(\eta_i^\mu)(\sigma_i(t))^2.
  \label{eq:begldef}
\end{equation}
This parameter can also be called fluctuation overlap since it can be
viewed as the retrieval overlap between the binary states $\sigma_i^2(t)$
and the patterns $\eta_i^\mu(t)$. It is, in general, independent of the
retrieval overlap $m^\mu(t)$. It gives rise to new states, the
so-called quadrupolar (or pattern-fluctuation retrieval) states with $m=0$
but $l \neq 0$. These states have a retrieval overlap zero but the activity
overlap is not, meaning that the active neurons $(\pm 1)$ coincide with
the active patterns but the signs are not correlated. Hence they carry
some retrieval information and  they might  be important in practical 
applications. In pattern recognition, e.g., looking at a black and white
picture on a grey background, these states would describe the situations
where the exact location of the picture with respect to the background 
is known but, the details of the picture itself are not focused. 
Furthermore, these  states might be helpful in
modelling such focusing problems discussed in the framework of
cognitive neuroscience \cite{SNK98}. 

The long-time behaviour of this network is governed by the following 
Hamiltonian \cite{DK00}, \cite{BV02}, precisely
obtained by optimizing the mutual information 
\begin{equation}
H = - \frac{1}{2} \sum_{i \neq j} J_{ij} \sigma_i \sigma_j
    - \frac{1}{2} \sum_{i \neq j} K_{ij} \sigma_i^2 \sigma_j^2  
    \label{ha1} \, .
\end{equation}
Since we want to compare this model with the $3$-Ising model and we
want to be able to change the relative importance of the two terms we 
rewrite the Hamiltonian as
\begin{equation}
H = - \frac{A}{2} \sum_{i \neq j} \tilde J_{ij} \sigma_i \sigma_j
    - \frac{B}{2} \sum_{i \neq j} \tilde K_{ij} \sigma_i^2 \sigma_j^2  
    \, ,
\end{equation}
with
\begin{equation}
   \tilde J_{ij}= a J_{ij} \, , \quad  \tilde K_{ij}= a(1-a) K_{ij} \, .
\end{equation}
For
\begin{equation} \label{BEGscaling}
A = \frac{1}{a}  \, , \quad  B = \frac{1} {a(1-a)}
\end{equation}
we trivially recover the model above.
When we now take $K_{ij} = b \delta_{ij}$ and $A=B=1$
we obtain the $3$-state Ising model (recall eq.(\ref{eq:s1})). 
Finally, we find back
the Hopfield model by taking first $B=0$ and then $a=1$, again with $A=1$.

For the ED and LF architectures we have to adapt the Hebbian learning rule
(\ref{beghebb}) analogously as in the $Q$-Ising model. For the ED case
both Hebbian weights are multiplied with the factor $c_{ij}N/C$, where
we recall that $c_{ij}$ is a random variable assuming values $0$ and $1$
with mean $C \approx O(ln N/N)$. For the LF architecture we consider   
\begin{equation}
J_{ij}(t)= 
\frac{1}{a^2 N} \sum_{\mu=1}^{p} \xi^{{\mu}}_i(t+1)\xi^{{\mu}}_j(t),
  \,\,\,  K_{ij}(t)= 
\frac{1}{ N} 
     \sum_{\mu=1}^{p} \eta^{{\mu}}_i(t+1)\eta^{{\mu}}_j(t) .
  \label{3}
\end{equation}
We remark that an underlying assumption that leads to the BEG model and
that should be preserved in any implementation is that the dynamic
activity $q \approx a$, as far as the order of magnitude is concerned.

\subsection{Solving the dynamics} \label{genscbeg}
The discussion given in Section 2.2 on the correlations appearing for
the various architectures remains valid for this model. Furthermore, the
development of the recursive scheme presented  there can be followed
in order to study the time evolution of the distribution of
the local fields $h_i(t)$ and $\theta_i(t)$. This allows one to write down
recursion relations determining the full time evolution of the order 
parameters (\ref{eq:begmdef})-(\ref{eq:begldef}) of the  model.

Since the method has been explained already in some detail in Section 2.2
and the explicit analysis is even more
technical we do not write it out here. For the FC architecture we 
refer to \cite{BBS} for the treatment at zero
temperature and to \cite{BBSV} for an extension to arbitrary 
temperatures. The final results are two recursion relations
of the type studied in Section 2.2.3, one for $h_i(t)$ and one for 
$\theta_i(t)$. 

Also for the BEG network the first few time steps of its evolution have 
been worked out analytically and have been compared with numerical 
simulations for systems up to $N=7000$ neurons averaged over $500$ runs.   
As an illustration we refer to fig.~4 left
presenting the order parameter $l$ as a function of $\alpha$ 
for uniform patterns and initial conditions $m_0=l_0=0.6, q_0=0.5$ and 
$T=0.5$. 
We remark that the maximal capacity for this system is
$\alpha_{c}\simeq 0.06$ (\cite{BV03}). We then learn that the
first time steps agree very well and do give a reasonable estimate
of the critical
capacity. 
\begin{figure}[ht] 
\centering\includegraphics[angle=270,width=.49\textwidth]{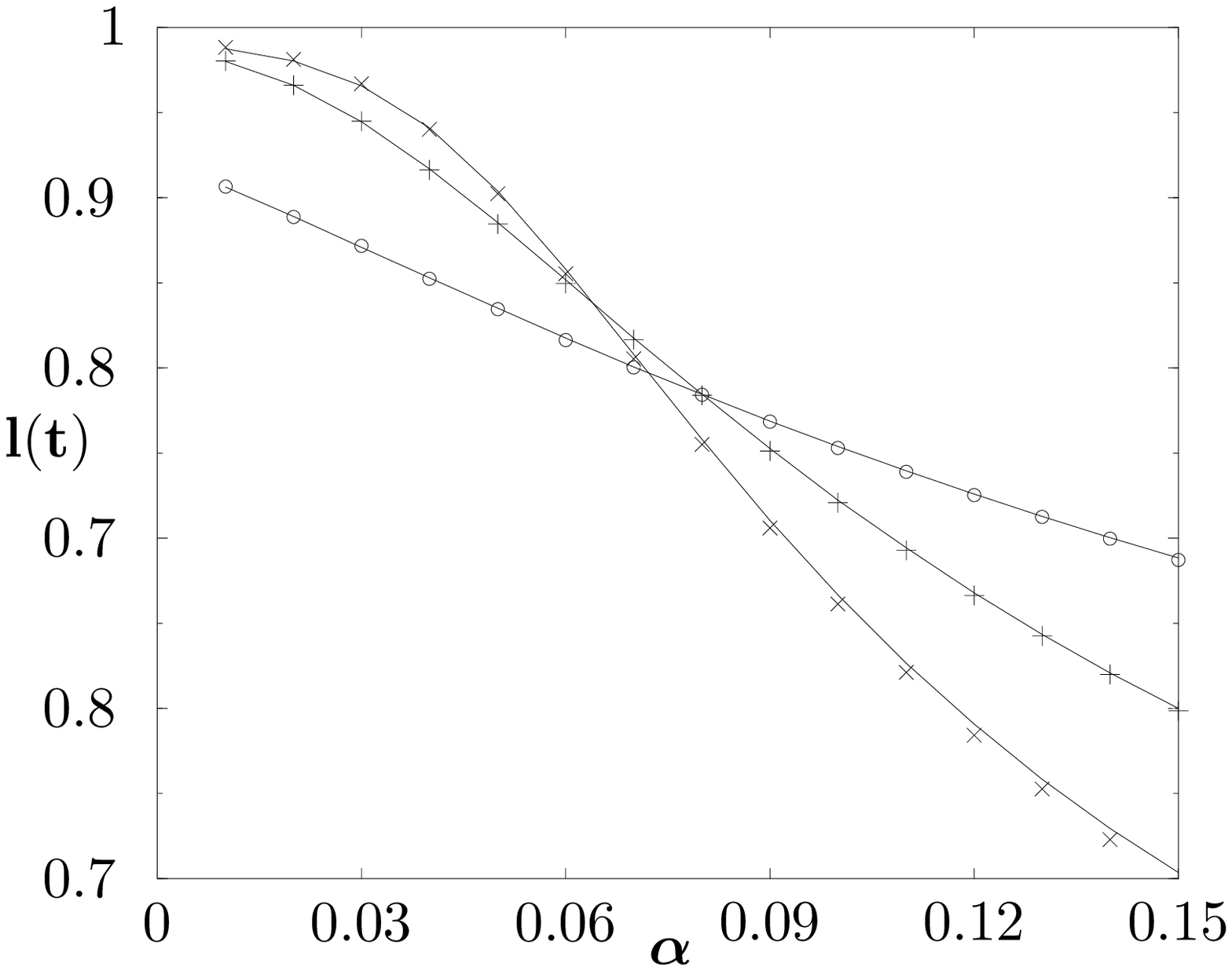}
\hfill
\centering\includegraphics[angle=270,width=.49\textwidth]{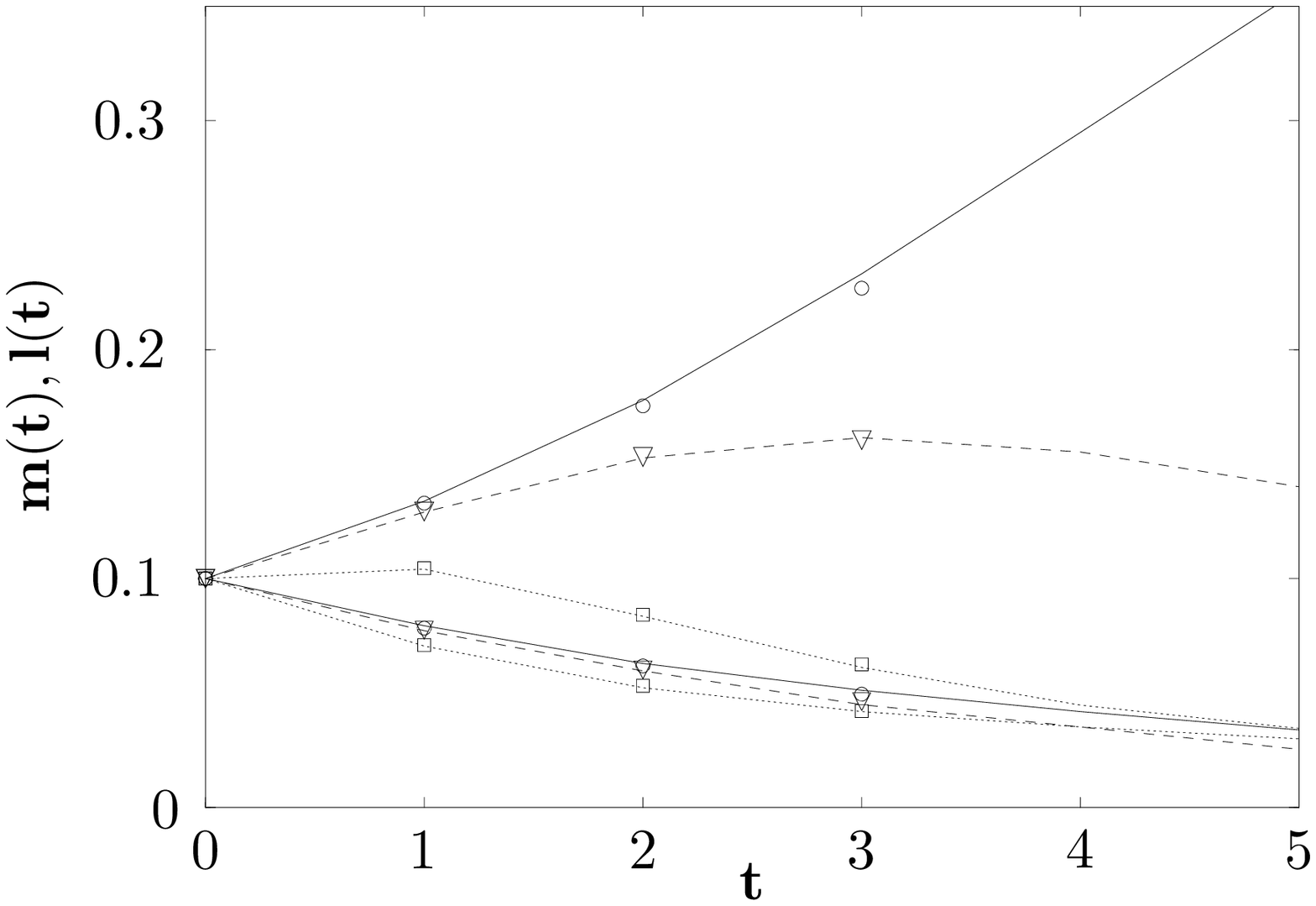} 
 \caption{\small The BEG model on a FC architecture with uniform patterns. 
 Left: Order parameter $l(t)$
as a function of the capacity $\alpha$ for the first three time
steps with initial conditions $m_0=l_0=0.6, q_0=0.5$
at $T=0.5$. Theoretical results (solid lines) versus simulations
(time $1,2$ and $3$ represented by a circle, a plus respectively
a times symbol) are shown.
Right: Order parameters $m(t)$ (bottom three lines) and $l(t)$ (top
three lines)  as a function of time with initial conditions 
$m_0=l_0=0.1, q_0=0.5$ at $T=1.1$ and several values of $\alpha$. 
Theoretical results (open symbols) versus simulations (full lines for 
$\alpha=0.001$, dashed lines for $\alpha=0.01$ and dotted lines for 
$\alpha=0.1$).}
\end{figure}

In fig.~4 we examine the order parameters $m$ and $l$ in the
quadrupolar phase ($m=0,l>0$) versus the paramagnetic phase
($m=0,l=0$), for several values of $\alpha$. We see that a few
time steps do give us already the characteristic behaviour. When
time increases $m$ decreases while $l$ differentiates between
the phases, as is seen in the theoretical results as well as in
the simulations. For the quadrupolar phase ($\alpha=0.001$) $l$
increases, deep inside the paramagnetic phase ($\alpha=0.1$) $l$
decreases, while in the intermediate region 
($\alpha=0.01$) the rate of increase of $l$ quickly diminishes and
$l$ itself goes to zero. 

At this point, we remark that there is a small but visible
discrepancy between the theory and simulations especially in
$l(3)$. It is of the order $O(10^{-3})$ and attributed to finite-size
effects. This,
and the fact that the signal-to-noise approach does not give a closed
form solution of the dynamics, has been a motivation  to look at the 
generating functional
approach to solve this dynamics. An extensive report is beyond the scope
of the present review. Essentially it turns out \cite{BBSV} that 
beyond the third time step of the dynamics the signal-to-noise analysis 
as used above is not entirely correct for
those parameters of the system corresponding to spin-glass behaviour.
The reason is that the technical assumption after eq. (\ref{eq:w}) is 
not valid in the spin-glass region but it seems to have little
effect in most of the retrieval region of the networks \cite{BBSV} 
(and \cite{BBVp} for full details in the simpler case of the Hopfield 
model).

To confirm this some further numerical experiments have been done for 
different values of the model parameters comparing this limiting normal
distribution  (recall eq. (\ref{eq:w}))
with simulations for different time steps. A comparison  for time steps
$t=2$ and $t=9$ for systems with
$N=2000$ neurons averaged over $250$ runs for the initial conditions
$m_0=l_0=0.6, q_0=0.5, a=2/3$ and temperature $T=0.2$ as a function
of $\alpha$  shows that in the retrieval region ($\alpha_c < 0.086$)
the simulation results coincide quite well with the limiting
distribution, while in the spin-glass region, certainly from
$\alpha \sim 0.11$ onwards, the results for $t=9$ start diverting
systematically \cite{BBSV}.  
We remark that the signal to noise approach can be used correctly by 
refining that technical
assumption allowing for the inclusion of all feedback correlations
\cite{BBVp}.

Concerning the other architectures we mention again that the AED and LF
models can be solved exactly \cite{DK00}, \cite{BDEKT},  
\cite{BET}, \cite{DKTE}, and we study the stationary
limit in the next subsection. Resulsts on the BEG model with variable
dilution, hence, including SED  can be found in \cite{V}.

\subsection{Thermodynamic and retrieval properties} \label{sec:fixpbeg}
Stationary results for the AED and LF architectures are obtained
immediately through the dynamical approach discussed in the previous
Section 3.2. 

The stationary states of the AED network dynamics  are
shown in Fig.5, for a typical activity of $a=0.8$ and $q\sim a$. The
pattern activity is chosen somewhat larger than $a=2/3$ (uniform
patterns) since for finite loading $\alpha=0$ it is easy to find out
that the quadrupolar state only exists for $a \geq 0.698$. In 
addition to the retrieval and quadrupolar phases, $R(m \neq 0,l \neq 0)$
and $Q(m=0,l \neq 0)$, there is a self-sustained activity phase 
$S(m=0,l=0)$, also referred to as the zero phase $Z$ \cite{DK00},
\cite{DKTE}. 
We remark that the saddle-points have one-dimensional
basins of attraction with attractor directions along $l$, either
towards $l^{*}\neq 0$ or to $l^{*}=0$ and repeller
directions along $m$ away from $m=0$. Furthermore, at the boundary of 
the maximal storage capacity $\alpha_{c}$,  both overlaps, $m$ and
$l$, disappear.

A similar behaviour appears for other big values of the
pattern activity $a$,
whereas for small $a$ there are only $R$ and $S$ phases. The
reason for a low-$T$ retrieval phase and the absence of a $Q$
phase is that a finite $T$ is needed for the active neurons ($\pm
1$) to coincide with the active patterns but with uncorrelated
signs, such that $m=0$. 
\begin{figure}[h]
\begin{center}
\includegraphics[angle=270,width=.49\textwidth]{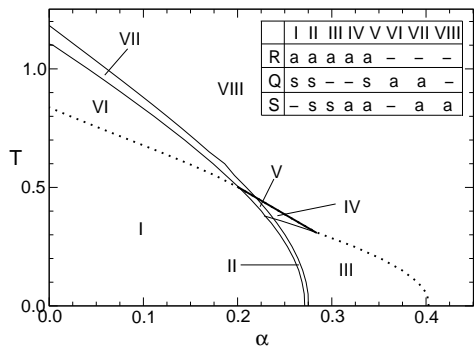}
\end{center}
\caption{\small  The $(T,\alpha)$ phase diagram for the AED
 BEG network with pattern activity $a=0.8$. Full (dotted) thin lines 
 denote discontinuous (continuous) transitions, thick lines denote the
boundary of the $R$ phase. The lines at the most right yield the
maximal storage capacity. The structure of the retrieval dynamics is
explained: a denotes an attractor, s a saddle point, r a repellor.}
\end{figure}

Again a lot of detailed results are available. The most important ones
can be summarized as follows. Above the threshold 
$(\alpha,T) =(0.22,0.45)$ a stable $Q$ phase 
starts to appear. For $T$ below that threshold  $m$
and $l$ remain finite together, in a behaviour characteristic for
retrieval, up to the maximal $\alpha_{c}$.
In this regime the fluctuation overlap does not yield anything
essentially new that is not contained in the retrieval overlap. In
contrast, above the threshold,  $m$ disappears first
with increasing $\alpha$ leaving a finite $l\neq 0$ up to a bigger
$\alpha_{c}$. Hence, first $T$ and then $\alpha$ have to become
large enough for the $Q$ states to appear. Note that the
fluctuation overlap carries a finite information even with $m=0$
in the $Q$ phase. Thus, 
although the information transmitted by the network is mainly in the 
retrieval phase, there is also some information due to the $Q$ phase.

For small $\alpha$, the fluctuation overlap
``drives'' a vanishingly small initial retrieval overlap, meaning almost
no recognition of a given pattern by the network, into an
asymptotic state with finite recognition. 
This is in contrast with the results for other three-state networks 
where first the overlap $m(t)$ becomes non-zero: $m(t)$ drives $l(t)$. 
Furthermore, with a vanishing initial $m_{0}$, the states 
of the network  pass through the vicinity of a saddle 
point $Q$, with a finite fluctuation overlap $l$ and still a
vanishing retrieval overlap at small or intermediate times, 
giving some  plateaus in $q$, $l$ and the information content. It is 
only in passing beyond those plateaus, which may take a rather
long time, that the states attain the asymptotic behaviour of the 
retrieval phase. 

In general, the basins of attraction for retrieval 
and the information content are larger in the BEG network than in other 
three-state networks. These results for
the dynamics and the stationary states are confirmed by flow diagrams
\cite{BDEKT}, \cite{DKTE}.

For the LF architecture some typical phase diagrams are shown in Fig.~6.
\begin{figure}
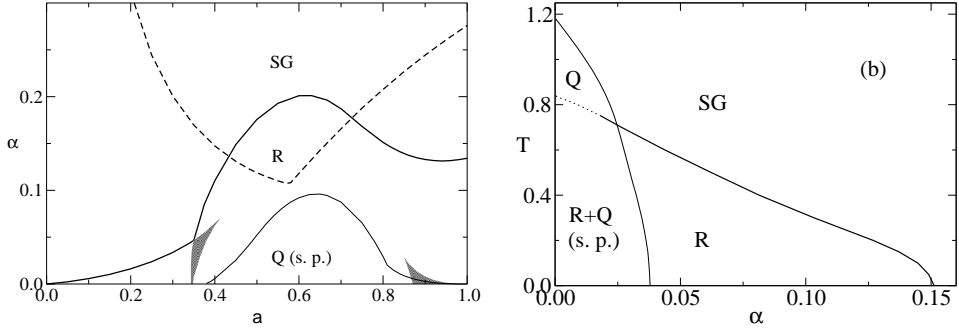

\rotatebox{270}{
\includegraphics[width=.34\textwidth]{novafig9.eps}}
\hfill
\rotatebox{270}{
\includegraphics[width=.34\textwidth]{novafig10.eps}}
\caption{\small The $(\alpha,a)$, $T=0$ (left) and $(T,\alpha)$, $a=0.8$ 
(right)
 phase diagram for the LF BEG network. There is a $SG$ solution 
everywhere. There is a stable $R$  phase (a second one with a smaller
overlap in the shaded area)  below the thick full  line. Left: The 
$Q$ state
appears as a saddle-point below the thin full line. The thick dashed
line shows the retrieval phase boundary for the optimal LF $Q=3$-Ising 
model. Right:  There is a  stable (saddle-point) $Q$ phase above (below)
the thick (dotted) line, ending discontinuously at the full thin line. }
\end{figure}
We first remark that we need to introduce two further variables in the
derivation of the LF recurrence relations for the variances of the two
noises i.e. 
\begin{equation}
q_1(t)=\left\langle \left\langle\sigma_i(t) 
\right\rangle_{\beta}^2\right\rangle_{\{\xi^\mu\}} \,,
\quad 
p_1(t)=\left\langle \left\langle\sigma_i^2(t) 
\right\rangle_{\beta}^2\right\rangle_{\{\xi^\mu\}}. 
\end{equation}
The possible phases are then $R(m>0,l>0,q_1>0,q_0>0)$, 
$Q(m=0,l>0,q_1>0,q_0>0)$ and $SG(m=0,l=0,q_1>0,q_0>0)$.
From Fig.~6 we notice that a stable $Q$ phase  only appears for 
sufficiently large $T$ and large $a$, a feature already seen
for the AED architecture. Thus, as $T$ increases, the useful performance 
of the network goes over from the retrieval to the pattern-fluctuation 
retrieval phase. Furthermore, we
see that for intermediate activity $a\in (0.435,0.727)$ the
LF BEG network has a bigger maximal storage capacity than the optimal 
LF Ising network \cite{BSV}, optimal in the sense  that the adjustable 
threshold parameter was chosen to optimize the storage capacity $\alpha$.
The same has been found for the information content. At larger activity, 
$a=0.8$ say,  the BEG and Ising networks compete for better performance 
at intermediate or larger $\alpha$ values. 
Moreover, as in the AED architecture, the
flows to the stable solutions are considerably delayed by the
saddle points in the form of slow transients of the dynamics. 
Finally, a remarkable feature is the presence of quite
large basins of attraction either to the stable $R$ state or to
the stable $Q$ state, even for the fairly high $T$ (and small
$\alpha$). Also, not surprisingly, one finds a much
smaller basin of attraction to the $SG$ states.  Similar features have
also been found in the dynamics of the AED network
except for the $SG$ states, which are absent in that case. 
 
For the symmetric architectures we restrict ourselves here to the FC
one. Results on the architecture with variable dilution can be found in
\cite{V}. We apply directly the standard replica technique
in order to calculate the free energy of the model.  Within the 
replica-symmetry approximation and  for
a finite number, $s$, of condensed patterns, we obtain
\begin{eqnarray} \label{free_energy}
 f(\beta) &=&  \frac{1}{2} \sum_{\mu=1}^s
       \left( a A\, m_\mu^2 + a(1-a) B \, l_\mu^2
       \right)
  + \frac{\alpha}{2\beta} \log (1-\chi)
  + \frac{\alpha}{2\beta} \log (1-\phi) \nonumber \\
   &+& \frac{\alpha}{2\beta} \frac{\chi}{1-\chi}
   + \frac{\alpha}{2\beta} \frac{\phi}{1-\phi}
  + \frac{\alpha}{2} \frac{A q_1 \chi}{(1-\chi)^2}
  + \frac{\alpha}{2} \frac{B p_1 \phi}{(1-\phi)^2} \nonumber \\
   &-& \frac{1}{\beta} \left\langle \int Ds Dt \ln
       \mbox{Tr}_\sigma
       \exp \left ( \beta \tilde H \right )
    \right\rangle_{\{\xi^\mu\}} \ ,
\end{eqnarray}
with the effective Hamiltonian $\tilde H$ given by
\begin{eqnarray}
\tilde H  &=&  A \sigma
 \left[
   \sum_\mu m_\mu \xi^\mu
        + \sqrt{\alpha r}s
 \right]
+ B \sigma^2
 \left[
   \sum_\mu l_\mu \eta^\mu + \sqrt{\alpha u} t
 \right] \nonumber\\
&+& \frac{\alpha}{2} \frac{  A \, \chi}{1-\chi} \, \sigma^2
+ \frac{\alpha}{2} \frac{  B \, \phi}{1-\phi} \, \sigma^2
\ .
\end{eqnarray}
Here
\begin{equation}
\chi = A \beta ( q_0 - q_1 )\, , \quad  \phi = B \beta ( p_0 - p_1)\, ,
   \quad r = \frac{q_1}{(1-\chi)^2}\, ,
   \quad u = \frac{p_1}{(1-\phi)^2} \, .
\end{equation}
In these expressions the relevant order parameters are
\begin{eqnarray}
m_\mu & = &
\frac{1}{a}
\left\langle  \xi^\mu \int \!  Ds Dt \
  \left\langle \sigma \right\rangle_\beta
\right\rangle_{\{\xi^\mu\}}
\ , \label{sp:m}
\\
l_\mu & = &
\frac{1}{a(1-a)}
\left\langle \eta^\mu \int \! Ds Dt \  \left\langle \sigma^2
\right\rangle_\beta
\right\rangle_{\{\xi^\mu\}}
\ , \label{sp:l}
\\
q_0 &  = &p_0 =
\left\langle
\int \! Ds Dt \  \left\langle \sigma^2 \right\rangle_\beta
\right\rangle_{\{\xi^\mu\}}
\ , \label{sp:q_0}
\\
q_1 & = &
\left\langle
\int \! Ds Dt \  {\left\langle \sigma
\right\rangle}_\beta^2
\right\rangle_{\{\xi^\mu\}}
\ , \label{sp:q_1}
\\
p_1 & = &
\left\langle
\int \! Ds Dt \  {\left\langle
\sigma^2 \right\rangle}^2_\beta
\right\rangle_{\{\xi^\mu\}}
\ , \label{sp:p_1}
\end{eqnarray}
where $\left\langle \cdot \right\rangle_\beta$ represents the thermal 
average with respect to $\tilde H$. As usual we take only one condensed 
pattern such that the index $\mu$ can be dropped.
The parameters $q_1$ and $p_1$ are the
Edwards-Anderson order parameters with their conjugate variables $r$
respectively $u$. Finally, $\chi$ and $\phi$ are the susceptibilities
proportional to the fluctuation of the $m$ overlap, respectively $l$
overlap. All these parameters are the stationary limits of the
corresponding parameters considered in the dynamics for arbitrary
temperatures.
We remark that the trace over the neurons and the average over the
patterns can be performed explicitly. The resulting expressions  are
written down in \cite{BV03} and have been solved numerically.
\begin{figure}[ht]
\centering\includegraphics[width=.49\textwidth]{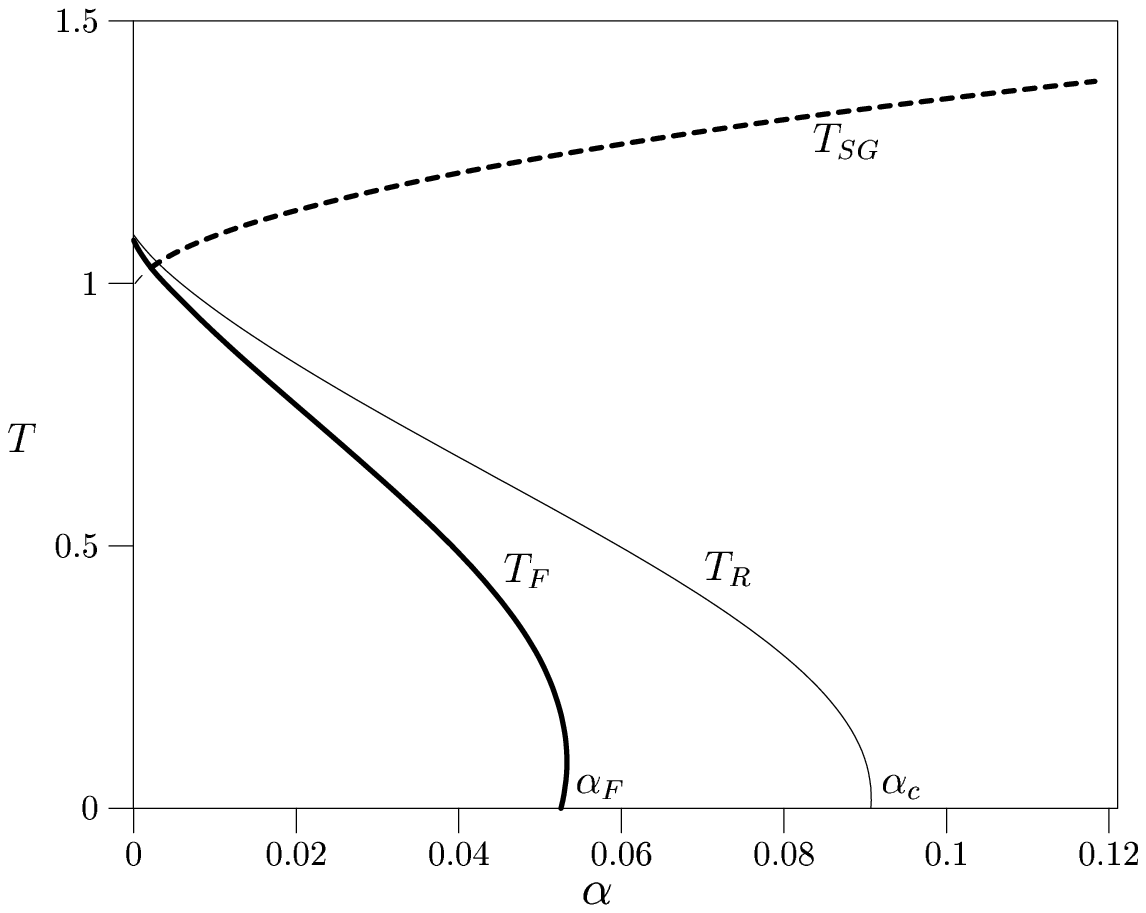}
 \hfill  
\centering\includegraphics[width=.49\textwidth]{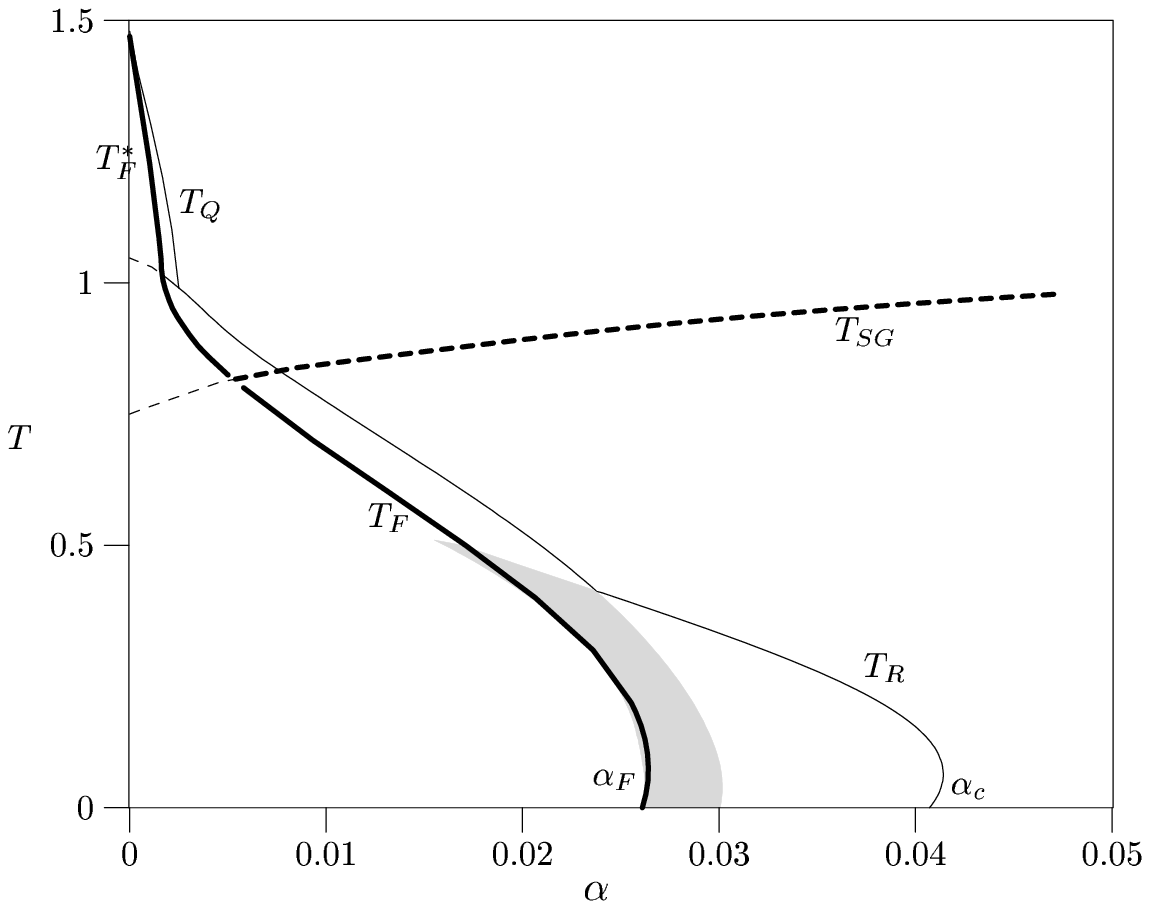}
 \caption{\small
    The BEG $(\alpha-T)$ phase diagram for $a=2/3$ (left) and $a=0.8$
    (right). Dashed lines correspond to continuous transitions,
while full lines correspond to discontinuous transitions (in all order
parameters). Below the line $T_R$ retrieval states occur.  The curve
$T_F$ represents the thermodynamic transition (shown as a thick line)
between retrieval states and spin-glass states. The line $T_{SG}$ 
denotes the transition from the spin-glass to the paramagnetic phase.
Below the line $T_Q$ quadrupolar states exist and below the line $T_F^*$
they are global minima. In the shaded region two retrieval states
coexist.  }
\end{figure}

From fig.~7 (left) for uniform patterns and a comparison with the results
for the FC $Q=3$-Ising model (cfr. fig~2 right) one
 learns that  $\alpha_c=0.091$ is almost double of the
maximal capacity for the latter, $\alpha_c=0.046$, in the case of an 
optimal choice for the gain parameter $b$, i.e. $b=1/2$.  Compared with
the
Hopfield  model, one sees that $\alpha_c$ is smaller in the BEG model,
$0.091$ versus $0.13$, but $\alpha_F$ is larger, $0.053$ versus $0.051$.
So a bigger number of the retrieval states in the BEG network are global
minima of the free energy. Finally, one also notices that
the critical curves $T_{SG}$ and $T_R$ end in different temperature
points at $\alpha=0$ giving rise to a `crossover' region for small
$\alpha$ as it typically occurs in other multi-state models, e.g., the 
Potts model \cite{Kanter}, \cite{BDH92} and the Askin-Teller model 
\cite{BK99}. This  is related
with the fact that for $\alpha=0$ these models have a discontinuous
transition at $T_F$. In this crossover region the retrieval states
(global minima below $T_F$) and the paramagnetic states (local minima
below $T_F$) coexist.

As in the AED and LF architecture the quadrupolar phase is situated in 
the high temperature region and we
can understand the physics behind it in the following way.  The spin-glass
order parameter $q_1$ is zero, meaning that the $\pm 1$ spins are not
frozen and as a consequence $m$ can be zero.  The fact that $l$ is not
zero practically means that the spins can flip freely between $\pm 1$
but the probability that they jump to 0 or vice versa becomes very small. 
This effect arises from $a>1/2$
onwards when the ratio between the second and the first term in the 
Hamiltonian starts increasing as $(1-a)^{-1}$. It implies that the
information content of the system is non-zero in this phase.

Finally, we recall that for $\gamma \equiv a(1-a)B=1$ 
(cfr. eq. (\ref{ha1})--(\ref{BEGscaling})), we
recover the BEG neural network as studied above. However, it
turns out that the maximum in the capacity is located at
$\gamma=0.712$ with a corresponding value of $\alpha_c = 0.096$.
A reason for this is the approximation $q_0 \sim a$ made in order to 
get the mean-field Hamiltonian.  The mutual information of
the network is optimized under this assumption but, in general, it may
not be completely realized in a specific model.
Furthermore, the fact that replica-symmetry breaking may be bigger for 
larger $\alpha$, as is also indicated by the zero-temperature entropy
calculation, could be an extra reason for this. For more details we
refer to the literature mentioned above.

\section{The Gardner capacity of multi-state models} \label{sec:crit}
In the previous Section it has been found that the capacity and basin of
attraction of the BEG network have been enlarged in comparison with those
of other three-state networks. The models considered all have
Hebbian-type learning rules. 
A natural question is then whether these improved retrieval quality
aspects are restricted to the use of the Hebb rule or whether they are 
an intrinsic property of the BEG model. Therefore, we want to answer the
following: given the set of $p$ patterns specified above, is
there a network (the best possible network of the BEG-type) which has
these patterns as fixed points of the deterministic form of the dynamics
(\ref{eq:beggain})?  

In order to do so we consider the perceptron architecture (N
inputs with couplings $J_j$ and $K_j$ and 1 output) and we say that a 
given pattern, $\xi_i^\mu, i=1, \ldots, N$,  is stored if there exists a 
corresponding output $\xi_0^\mu$
\begin{equation}
\xi ^\mu_0=g(h ^\mu,\theta ^\mu) \label{metastablecondition}
\end{equation}
with
\begin{equation}
h ^\mu=\frac{1}{\sqrt{N}}\sum_{j=1}^N J_j\xi_j^\mu
\quad\quad 
\theta^\mu =\frac{1}{\sqrt{N}}\sum_{j=1}^N K_j (\xi_j^\mu)^2 
 \, ,
\label{cond2}
\end{equation}
and $\{{\bf J},{\bf  K}\} \equiv \{J_j, K_j\}$ denoting the
configurations in the space of interactions. The factor $N^{-1/2}$ is
introduced to have the weights $J_j$ and $K_j$ of order unity (spherical
constraint).

The aim is then to determine the maximal number of patterns, $p$, that can
 be
stored in the perceptron, in other words to find the maximal value of the 
loading $\alpha=p/N$  for which couplings satisfying
(\ref{metastablecondition})-(\ref{cond2}) can still be found. Following a
Gardner-type analysis \cite{Ga} the fundamental quantity that we want to
calculate is then the volume fraction of weight space  given by
\begin{equation}
V=\int d {\bf J} d{\bf K}\rho({\bf J},{\bf
K})\prod_{\mu=1}^p\chi_{\xi^\mu_0 }(h ^\mu,\theta^\mu ;\kappa)
\label{accesiblevolume}
\end{equation}
with the characteristic function
\begin{eqnarray}
 \chi_{\xi^\mu_0 }(h ^\mu,\theta^\mu ;\kappa) 
  &&=\delta_{\xi ^\mu_0,g(h ^\mu,\theta ^\mu)} \nonumber \\
  &&=(\xi^\mu_0)^2 \Theta(|h^\mu|+\theta ^\mu-\kappa)
              \Theta(\xi^\mu_0  h^\mu -\kappa) \nonumber \\
  &&+(1-(\xi^\mu_0 )^2)\Theta(-|h ^\mu|-\theta ^\mu-\kappa) 
\label{character}
\end{eqnarray}
where $\kappa$ is the imbedding stability parameter measuring the size of
the basin of attraction for the $\mu$-th pattern and  $\rho({\bf J},{\bf
K})$ is the following normalization factor assuming  spherical
constraints for the couplings
\begin{equation}
\rho({\bf J},{\bf K})=\frac{\delta({\bf J}\cdot{\bf J}-N)\delta({\bf
K}\cdot{\bf K}-N)}{\int_{-\infty}^\infty  d {\bf J} d{\bf K}\delta({\bf
J}\cdot{\bf J}-N)\delta({\bf K}\cdot{\bf K}-N)} \, .
 \label{normalization}
\end{equation}
In order to perform the average over the disorder in the input patterns
and the corresponding output we employ the replica technique to evaluate
the entropy per site
\begin{equation}
v =\lim_{N\to\infty}\frac{1}{N}\left\langle \left\langle
\ln V \right\rangle \right\rangle\label{entropy}
\end{equation}
where  $\left\langle \left\langle\cdots\right\rangle \right\rangle$ denotes
 an average over the statistics of
inputs $\{\xi_j^\mu\}$ and outputs $\{\xi_0^\mu\}$, recalling
(\ref{begprobability}).

In the replica approach the entropy per site $v$ is computed via the
expression
\begin{equation}
v=\lim_{N\to\infty}\lim_{n\to 0}\frac{1}{nN}
    \big(\left\langle\left\langle
               V^n\right\rangle\right\rangle-1\big)
   =\lim_{N\to\infty}\lim_{n\to 0}\frac{1}{nN}\ln \left\langle\left\langle
                  V^n\right\rangle\right\rangle
\end{equation}
where $V^n$ is the $n$-times replicated fractional volume
\begin{eqnarray}
\left\langle\left\langle V^n\right\rangle\right\rangle \propto 
  \int \Big[\prod_{\alpha=1}^n d {\bf J}^\alpha d {\bf K}^\alpha 
    \delta\Big({\bf J}^\alpha\cdot{\bf J}^\alpha-N\Big)
    \delta\Big({\bf K}^\alpha\cdot{\bf K}^\alpha-N\Big) \Big]
       \nonumber \\
       \times
 \left\langle\left\langle
     \prod_{\alpha=1}^n\prod_{\mu=1}^p
     \chi_{\xi^\mu_0}(h ^\alpha_\mu,\theta ^\alpha_\mu;\kappa)
 \right\rangle\right\rangle
\label{fracvol}
\end{eqnarray}
whereby we can forget, since the couplings are continuous, about constant
terms such as the  denominator in (\ref{normalization}). The
replica-symmetric calculation then proceeds in a standard way, although 
the technical details are much more complicated, and an analytic formula
can be obtained \cite{BPS}.

Comparing with analogous discussions in the literature for other 
three-state neuron perceptron models we recall that 
for $\kappa=0$ and uniform patterns  the $Q=3$ Ising perceptron can 
maximally reach an optimal capacity equal to $1.5$,  depending on the 
separation between the plateaus of the gain function  (see \cite{MKB},
 \cite{BM} for the precise details) and the  $Q=3$ clock and Potts model
both reach an optimal capacity of $2.40$ \cite{GK94}, \cite{GBK} while the
value for the BEG perceptron found here is $2.24$. Here we have to recall 
that the $Q=3$ Ising perceptron and the  BEG perceptron have the same 
topology structure in the neurons, whereas the $Q=3$ clock and Potts models
have a different topology, as explained in the Introduction. 
Since, in general, perceptrons turn out to be very useful models in 
connection with learning and generalization this is an interesting 
observation. 

The stability of the replica-symmetric solution has been studied by 
generalizing the de Almeida-Thouless analysis and deriving an analytic 
expression for the {\it two} replicon eigenvalues that play a role in 
the Gardner limit. Breaking only occurs for small activities and very
small imbedding constants, $\kappa < 0.0061$. This is consistent with
the stability results found for the $Q=3$ Ising perceptrons. 

These results strenghten the idea that the better retrieval properties 
found for the BEG model in comparison with the $Q=3$ 
Ising model are not restricted to the specific Hebb rule but are 
intrinsic to the model.

\section{Concluding remarks} \label{sec:con}
In this overview we have studied the dynamics and retrieval properties
of multi-state neural networks based upon spin-glass models. In
particular, we have first discussed the $Q$-Ising model and the
Blume-Emery-Griffiths model with various architectures and Hebbian-type
learning rules. The methods used are the signal-to-noise analysis and the
thermodynamic mean-field replica technique. Then, the Gardner 
optimal capacity for these models has been considered. 

A number of
detailed results have been outlined  in order to compare the properties of 
the different networks and architectures. The Blume-Emery-Griffiths model, 
obtained by maximizing the mutual information content of networks with 
scalar valued three-state neurons, shows improved retrieval properties in 
comparison with the $Q=3$-Ising model.

\section*{Acknowledgments}

This work has been supported in part by the Fund for Scientific
Research, Flanders-Begium.  The author is indebted to
S.~Amari, J. Busquets Blanco, D. Carlucci,  D. Dominguez, R. Erichsen Jr., 
J. Huyghebaert, G. Jongen, E. Korutcheva,  P. Kozlowski, R.~K\"uhn, 
I. P\'erez Castillo, H. Rieger, 
G.M. Shim, W.K. Theumann,  J. van Mourik, T. Verbeiren, B.Vinck, K.Y.M.
Wong and V.~Zagrebnov for pleasant collaborations on some of these 
subjects during  previous years. He especially
thanks T. Verbeiren for discussions about the present text.

\end{document}